\documentclass[%
amsmath,amssymb,linenumber,
twocolumn,
superscriptaddress,
amsmath,amssymb,
prl,
longbibliography
]{revtex4-1}

\usepackage{graphicx}
\usepackage{dcolumn}
\usepackage{bm}
\usepackage{multirow}

\usepackage{color}

\begin{document}


\renewcommand{\vec}[1]{\ensuremath{\boldsymbol{#1}}}
\newcommand{\mat}[1]{\ensuremath{\boldsymbol{#1}}}
\newcommand{\Transpose}{\ensuremath{^T}} 

\newcommand{\PP}{\ensuremath{\mathds{P}}}
\newcommand{\RR}{\ensuremath{\mathds{R}}}
\newcommand{\NN}{\ensuremath{\mathds{N}}}

\newcommand{\InnerProd}[3][]{\ensuremath{%
\ifthenelse{\equal{#1}{}}{\left<{#2},{#3}\right>}{%
\ifthenelse{\equal{#1}{1}}{\bigl<{#2},{#3}\bigr>}{%
\ifthenelse{\equal{#1}{2}}{\Bigl<{#2},{#3}\Bigr>}{%
\ifthenelse{\equal{#1}{3}}{\biggl<{#2},{#3}\biggr>}{%
\ifthenelse{\equal{#1}{4}}{\Biggl<{#2},{#3}\Biggr>}{%
\typeout{Warning: Invalid size specifier #1 for \InnerProd.}\InnerProd{#2}{#3}}}}}}}}

\newcommand{\Set}[3][]{\ensuremath{%
\ifthenelse{\equal{#1}{}}{\left\{#2\:\left|\:#3\vphantom{#2}\right.\right\}}{%
\ifthenelse{\equal{#1}{1}}{\bigl\{#2\:\bigl|\:#3\bigr.\bigr\}}{%
\ifthenelse{\equal{#1}{2}}{\Bigl\{#2\:\Bigl|\:#3\Bigr.\Bigr\}}{%
\ifthenelse{\equal{#1}{3}}{\biggl\{#2\:\biggl|\:#3\biggr.\biggr\}}{%
\ifthenelse{\equal{#1}{4}}{\Biggl\{#2\:\Biggl|\:#3\Biggr.\Biggr\}}{%
\Set{#2}{#3}}}}}}}}

\newcommand{\Norm}[2][]{\ensuremath{%
\ifthenelse{\equal{#1}{}}{\left\|{#2}\right\|}{%
\ifthenelse{\equal{#1}{1}}{\bigl\|{#2}\bigr\|}{%
\ifthenelse{\equal{#1}{2}}{\Bigl\|{#2}\Bigr\|}{%
\ifthenelse{\equal{#1}{3}}{\biggl\|{#2}\biggr\|}{%
\ifthenelse{\equal{#1}{4}}{\Biggl\|{#2}\Biggr\|}{%
\typeout{Warning: Invalid size specifier #1 for \Norm.}\Norm{#2}}}}}}}}

\newcommand{\AlgRef}[1]{Algorithm~\ref{#1}} 
\newcommand{\FigRef}[1]{Fig.~\ref{#1}} 
\newcommand{\FigRefTwo}[2]{Figs.~\ref{#1}, \ref{#2}}
\newcommand{\TabRef}[1]{Table~\ref{#1}} 
\newcommand{\EquRef}[1]{(\ref{#1})} 
\newcommand{\EquRefTwo}[2]{(\ref{#1},\ref{#2})}
\newcommand{\PageRef}[1]{p.\,\ref{#1}} 
\newcommand{\ExRef}[1]{example~\ref{#1}} 
\newcommand{\citeRef}[1]{ref.~\cite{#1}} 
\newcommand{\citeRefMany}[1]{refs.~\cite{#1}}

\newcommand{\NormalDistrSymbol}{\ensuremath{\mathcal{N}}}
\newcommand{\NormalDistr}[3][]{\ensuremath{%
\ifthenelse{\equal{#1}{}}{\NormalDistrSymbol\!\left({#2,#3}\right)}{%
\ifthenelse{\equal{#1}{0}}{\NormalDistrSymbol\!({#2,#3})}{%
\ifthenelse{\equal{#1}{1}}{\NormalDistrSymbol\!\bigl({#2,#3}\bigr)}{%
\ifthenelse{\equal{#1}{2}}{\NormalDistrSymbol\!\Bigl({#2,#3}\Bigr)}{%
\ifthenelse{\equal{#1}{3}}{\NormalDistrSymbol\!\biggl({#2,#3}\biggr)}{%
\ifthenelse{\equal{#1}{4}}{\NormalDistrSymbol\!\Biggl({#2,#3}\Biggr)}{%
\typeout{Warning: Invalid size specifier #1 for \NormalDistrSymbol.}\NormalDistrSymbol\!{#2,#3}}}}}}}}}

\newcommand{\Prob}[1]{\ensuremath{\PP\left({#1}\right)}}         
\newcommand{\ProbCond}[3][]{\ensuremath{%
\ifthenelse{\equal{#1}{}}{\PP\!\left(#2\left|\,#3\vphantom{#2}\right.\right)}{%
\ifthenelse{\equal{#1}{1}}{\PP\bigl(#2\,\bigl|\,#3\bigr.\bigr)}{%
\ifthenelse{\equal{#1}{2}}{\PP\Bigl(#2\,\Bigl|\,#3\Bigr.\Bigr)}{%
\ifthenelse{\equal{#1}{3}}{\PP\biggl(#2\,\biggl|\,#3\biggr.\biggr)}{%
\ifthenelse{\equal{#1}{4}}{\PP\Biggl(#2\,\Biggl|\,#3\Biggr.\Biggr)}{%
}}}}}}}

\newcommand{\CovarOp}{\ensuremath{\mathrm{covar}}}                  
\newcommand{\Covar}[2][]{\ensuremath{%
\ifthenelse{\equal{#1}{}}{\CovarOp\!\left({#2}\right)}{%
\ifthenelse{\equal{#1}{0}}{\CovarOp({#2})}{%
\ifthenelse{\equal{#1}{1}}{\CovarOp\bigl({#2}\bigr)}{%
\ifthenelse{\equal{#1}{2}}{\CovarOp\Bigl({#2}\Bigr)}{%
\ifthenelse{\equal{#1}{3}}{\CovarOp\biggl({#2}\biggr)}{%
\ifthenelse{\equal{#1}{4}}{\CovarOp\Biggl({#2}\Biggr)}{%
\typeout{Warning: Invalid size specifier #1 for \Covar.}\Covar{#2}}}}}}}}}

\newcommand{\Gradient}[1][]{\ensuremath{\mathbf{\nabla}\ifthenelse{\equal{#1}{}}{}{_{\!\vec{#1}\,}}}}

\newcommand{\Xdom}{\ensuremath{\mathcal{X}}}

\def\beq{\begin{equation}}
\def\eeq{\end{equation}}
\def\bea{\begin{eqnarray}}
\def\eea{\end{eqnarray}}
\def\fatR{{\bf R}}
\def\fatd{{\bf d}}
\def\fatr{{\bf r}}
\def\fatM{{\bf M}}
\def\fatR{{\bf R}}
\def\fatr{{\bf r}}
\def\fatM{{\bf M}}
\def\fatalpha{{\bm \alpha}}
\def\fatK{{\bf K}}
\def\fatI{{\bf I}}
\def\fatp{{\bf p}}
\def\ie{{\em i.e.~}}

\preprint{APS/123-QED}

\title[]{Machine Learning, Quantum Mechanics, and Chemical Compound Space}
\author{Raghunathan Ramakrishnan} 
\email{r.ramakrishnan@unibas.ch}
\affiliation{Institute of Physical Chemistry and National Center for Computational Design and Discovery of Novel  Materials (MARVEL), 
Department of Chemistry, University of Basel, Klingelbergstrasse~80, CH-4056 Basel, Switzerland}
\author{O. Anatole von Lilienfeld}
\email{anatole.vonlilienfeld@unibas.ch}
\affiliation{Institute of Physical Chemistry and National Center for Computational Design and Discovery of Novel  Materials (MARVEL), 
Department of Chemistry, University of Basel, Klingelbergstrasse~80, CH-4056 Basel, Switzerland}

\begin{abstract}
We review recent studies dealing with the generation of machine learning models of molecular and materials properties.
The models are trained and validated using standard quantum chemistry results obtained for organic molecules and
materials selected from chemical space at random.
\end{abstract}
    
\maketitle

\section{Introduction}
Over the last couple of years a number of machine learning (ML) studies have appeared  
which all have in common that quantum mechanical properties are being predicted based
on regression models defined in chemical compound space (CCS).
CCS refers to the combinatorial set of all compounds that could possibly be isolated and constructed
from all possible combinations and configurations of atoms, it is unfathomably huge.  
For example, in a 2004 {\em Nature Insight} issue, the number of small organic molecules, expected
to be stable, has been estimated to exceed 10$^{60}$~\cite{ChemicalSpace,Chemicalspace_Lipinski2004,Chemicalspace_Dobson2004}.
By contrast, current records held by the Chemical Abstract Services of the American Chemical Society
account for only $\sim$100 Million compounds characterized so far.
Conventionally, CCS has been a common theme in the experimental branches of chemistry, 
as well as in chem-informatics and bioinformatics. 
A review of the quantum mechanical first principles, or {\em ab initio}, view on CCS has recently been published by one of us~\cite{anatole-ijqc2013}.
The quantum mechanical framework is crucial for the unbiased exploration of CCS since it enables, 
at least in principle, the free variation of nuclear charges, atomic weights, atomic configurations, and electron-number.
These variables feature explicitly in the system's Hamiltonian. 
For most molecules the Born-Oppenheimer approximation is applicable in their relaxed geometry, 
rendering the calculation of solutions to the electronic Schr\"odinger equation the most time-consuming aspect.
Any {\em ab initio} approximation, be it density functional theory (DFT), post-Hartree-Fock (post-HF), 
or quantum Monte Carlo (QMC) based, can be used to that end~\cite{AbInitioDefinitionByKieronBurke}.
By virtue of the correspondence principle in quantum mechanics (QM)~\cite{MolecularElectronicStructureTheory}, 
and through the appropriate application of statistical mechanics (SM)~\cite{tuckerman_book_SM},
it should be clear that any observable can be calculated, or at least estimated, for any chemical species 
in CCS---as long as sufficient computational resources are available.
For select observables including energies, forces, or few other properties,
semi-empirical quantum chemistry methods or universal or reactive force-fields (FFs)~\cite{UFFRappeGoddard1992,ReaxFF2001,Grimme_QMDFF2014,Popelier_QCTFF2015} 
can also be used instead of QM, provided that they are sufficiently accurate and transferable.

Due to all the possible combinations of assembling many and various atoms 
the size of CCS scales formally with compound size as $\mathcal{O}(Z_{max}^{N_I})$ in an exponential manner.
Here, $Z$ refers to the nuclear charge of an atom ($Z$ is a natural number), 
and $Z_{max}$ corresponds to the maximal permissible nuclear charge in Mendeleev's table, typically $Z_{max} < 100$
(theoretical estimates exist though for $Z <$ 173~\cite{Pekka137}),
and $N_I$ corresponds to the number of nuclear sites. 
$N_I$  is a natural number and it defines the extension of 
a ``system'', and it can reach Avogadro number scale for assemblies of cells, or amorphous and disordered matter.
More rigorously defined within a Born-Oppenheimer view on QM, CCS is the set of combinations, 
for which the potential energy $E$ is minimal in real atomic positions $\{\fatR_I\}$, 
for given nuclear charges $\{Z_I\}$, and positive electron number. 
Within this framework, we typically do not account for temperature effects, isotopes, 
grand-canonical ensembles, environmental perturbations, external fields, relativistic effects or nuclear quantum effects. 
At higher temperatures, for example, the kinetic energy of the atoms can be too large to allow for
detectable life-times of molecular structures corresponding to shallow potential energy minima.
For a more in-depth discussion of stability in chemistry versus physics see also Ref.~\cite{RealismStability}.
The dimensionality of CCS corresponds then to 4$N_I+$3 because for each atom there are 
three spatial degrees of freedom in Cartesian space, one degree of freedom corresponding to nuclear charge. 
An additional degree of freedom specifies the system's handedness.
Finally, there are two electronic degrees of freedom: 
One representing the number of electrons ($N_e$), the other the electronic state (molecular term symbol).
Rotational and translational degrees of freedom can always be subtracted. 
While certainly huge and high-dimensional, it is important to note that CCS is by no means infinite. 
In fact, the number of possible nuclear charges is quite finite, 
and due to near-sightedness long-range contributions rarely 
exceed the range of 1 nm in neutral systems~\cite{KohnNearsightedness}.
Also, molecular charges are typically not far from zero, i.e. $N_e \sim \sum_I Z_I$.
Furthermore, it should also be obvious that not all chemical elements ``pair'' in terms of nearest neighbors.
As such, we consider it self-evident that there is a host of exclusion rules and redundancies
which result in a dramatic lowering of the effective dimensionality of CCS. 
And hence, seen as a QM property space spanned by combinations of atomic configurations and identities, 
it is natural to ask how interpolative regression methods perform when used 
to {\em infer} QM observables, rather than solving the conventional quantum chemistry approximations 
to Schr\"odinger's equation (SE)~\cite{JensenCompChem}.

Within materials and chem-informatics, ML and inductive reasoning are known for their use in so called
quantitative structure property (or activity) relationships (QSPR/QSAR).
Despite a long tradition of ML methods in pharmaceutical
applications~\cite{Kubinyi,SignatureFaulon2003,Golbraikh2003,WienerDescriptors,Klaus_EditorialChemicalSpace2011},
and many success stories for using them as filters in order to rapidly condense large molecular libraries~\cite{VirtualScreeningReview_2013},
their overall usefulness for molecular design has been questioned~\cite{SchneiderReview2010}.
In our opinion, these models fail to systematically generalize to larger and more diverse sets of molecules, compounds, or properties
because they lack a rigorous link to the underlying laws of QM and SM.
By contrast, within the realm of theoretical physical chemistry,
the fitting of functions to model physical relationships is well established and hardly controversial.
The use of regression methods has a long-standing track record of success in theoretical chemistry 
for rather obvious reasons: The cost of solving SE has been and still is substantial 
(due to polynomial scaling with number of electrons in the molecule), 
and the dimensionality of the movements of atoms is high, i.e.~directly proportional to the number of atoms in the system.
The issue becomes particularly stringent in the context of accurately fitting
potential energy surfaces (PES) to rapidly calculate spectroscopic properties, perform molecular dynamics (MD), 
or calculate phonons in solids. 
Fitting PES is a classic problem in molecular quantum chemistry, and has triggered substantial research ever since Herzberg's seminal contributions.
The studies of Wagner, Schatz and Bowman introduced the modern computing 
perspective in the 80s~\cite{WagnerSchatzBowman_PESfit1981,Schatz_PESreview1989}.
Neural network fits to PES started to appear with Sumpter and Noid in 1992~\cite{SumpterNoidNeuralNetworks1992},
followed by 
Lorenz, Gross and Scheffler in 2004~\cite{Neuralnetworks_Scheffler2004},
Manzhos and Carrington in 2006~\cite{NN_Tucker2006},
and Behler and Parrinello in 2007~\cite{Neuralnetworks_BehlerParrinello2007}, 
and others~\cite{PhysRevB.92.045131}. 
The latter contributions mostly use the resulting models for extended MD calculations.
Kernel based ML methods for fitting potentials have also been used for similar problems
already in 1996~\cite{Rabitz1996}, and more recently in 2010~\cite{bpkc2010}.
For more information on fitting force-fields using flexible and highly parametric functions,
rather than models inspired by physical models of interatomic interactions with minimal parameter sets, 
the reader is referred to the recent reviews in Refs.~\cite{Behler_NNreview2015,ManzhosCarrington_NNreview2015,RampiMLQMMM}. 

ML models trained across chemical space are significantly less common in physical chemistry.
Notable exceptions, apart from the papers we review here-within, include (in chronological order) 
attempts to use ML for the modeling of 
enthalpies of formation~\cite{NN4B3LYP_Chen2003},
density functionals~\cite{NN4B3LYP_Chen2004},
melting points~\cite{Bender_Melting2005},
basis-set effects~\cite{SVM4CBS_Lomakina2011},
reorganization energies~\cite{anatole-MilindDenis2011},
chemical reactivity~\cite{Baldi_Reactivity2012},
solubility~\cite{Baldi_Solubility2013},
polymer properties~\cite{ML4Polymers_Rampi2013},
crystal properties~\cite{GrossMLCrystals2014,ML4Crystals_Wolverton2014},
or frontier orbital eigenvalues~\cite{Alan_OLED2015}.
Automatic generation of ML models, trained across chemical space, of the expectation value of the quantum mechanical 
Hamiltonian of a realistic molecule, i.e.~the observable energy, 
has only been accomplished within the rigorous realm of physical chemistry~\cite{RuppPRL2012}.
More specifically, this seminal contribution demonstrated that one can also use ML instead of 
solving the SE within the Born-Oppenheimer approximation in order to find the potential energy of 
a system as the eigenvalue of the electronic Hamiltonian, $\hat{H}(\{Z_I, \fatR_I\}) \stackrel{\rm \Psi}{\longmapsto} E$, using its wavefunction $\Psi$ for a given set of nuclear charges $\{Z_I\}$ and atomic positions $\{\fatR_I\}$. 
In other words, it is possible to generate a regression model which 
(i) directly maps nuclear charges and positions to potential energy, $\{Z_I, \fatR_I\} \stackrel{\rm ML}{\longmapsto} E$,
and (ii) can be improved systematically through mere addition of training instances.
The electronic Hamiltonian $H$ for solving the non-relativistic time-independent SE within the Born-Oppenheimer approximation
and without external electromagnetic fields,
$H\Psi = E\Psi$, of {\em any} compound with a given charge, $Q = N_p - N_e$,
is uniquely determined by its external potential, $v(\fatr) = \sum_I Z_I/|\fatr-\fatR_I|$, i.e.~by $\{\fatR_I,Z_I\}$.
As such, relying on SE, as opposed to alternative interatomic or coarse-grained based {\em ad hoc} models, 
has the inherent advantage of universality and transferability---an aspect which is conserved in our ML models. 
This aspect is crucial for the usefulness of the ML models generated in the context of explorative studies of CCS, e.g.~when devising virtual compound design strategies. 
Throughout our work we strive to maintain and apply the same fundamental QM based rigor that 
also underlies the success of quantum chemistry for modeling molecules, liquids, and solids. 

\paragraph{Paradigm}
Our {\em Ansatz} to chemistry relies on combining conventional QM based "Big Data" generation with supervised learning.
The paradigm rests squarely on inductive reasoning, being applied in the rigorous context of supervised learning and QM.
Generally speaking, we believe that this ML approach can be applied successfully to the modeling of physical properties of chemicals if 
\begin{itemize}
\item[(i)] there is a rigorous cause and effect relationship connecting system to property, 
such as the quantum mechanical expectation value of a ground-state property being coupled to the system's Hamiltonian via its wavefunction;
\item[(ii)] the representation accounts for all relevant degrees of freedom necessary to reconstruct the system's Hamiltonian;
\item[(iii)] the query scenario is interpolative in nature, meaning that for any query the relevant degrees of freedom assume values in between (or very close to) training instances;
\item[(iv)] sufficient training data is available.
\end{itemize}

Here, we refer to ML as an inductive supervised-learning approach that does not require {\em any}
a priori knowledge about the functional relationship that is to be modeled, and that improves
as more training data is being added.
This approach corresponds to the mathematically rigorous implementation of inductive reasoning
in the spirit of Refs.~\cite{MueMikRaeTsuSch01,HasTibFri01,MLbook,RasmussenWilliams}:
Given sufficient examples of input and corresponding output variables
any function relating the two can be modeled through statistics, and subsequently be used to infer
solutions for new input variables---as long as they fall into the interpolating regime.
The amount of training data necessary to be considered ``sufficient'' depends on 
(a) the desired accuracy of the ML model, 
(b) its domain of applicability, 
and (c) its efficiency. 
Overfitting and transferability can be controlled through careful regularization 
and cross-validation procedures.
As such, it offers a mathematically rigorous way to circumvent the need to establish an assumed approximate physical model in terms of input-parameters or constraints.
In chemistry, the historical importance of inductive reasoning is considerable.
Examples include the discovery of Mendeleev's table, Hammett's equation~\cite{Hammett_cr1935,Hammett_jacs1937},
Pettifor's structure maps~\cite{StructureMaps_Pettifor1986}, or also
the Evans-Polanyi, Hammond's, or Sabatier's principles, 
currently popular as volcano plots used for computational heterogeneous catalyst design~\cite{ReviewCatalystNorskov}.
Philosophically speaking, even Newton's law and the postulates of quantum mechanics rely on inductive reasoning.
For a rigorous QM based ML approach, however, all inherent heuristic aspects should be rooted out in order to 
arrive at truly {\em transferable} model functions whose accuracy and applicability can 
be improved {\em systematically}, i.e.~through mere addition of more data.

We reiterate that in {\em any} data-driven model generation process, a large number of training examples 
are required to develop converged, well performing models.  The details of the origin are irrelevant to the model. 
If, for example, thousands of Lennard-Jones (LJ) potential energies are being provided, 
the ML model will attempt to model the LJ function. 
Since there is no noise in this function, the noise level in the ML model will be negligible.
If, by contrast, through some randomized process, sometimes a LJ energy, 
and sometimes a Morse-potential energy is provided as training data, the ML model must cope with the resulting noise. 
Note that training data could also have experimental origin.
While it is possible, of course, to train ML models on known functions, there is little merit in such models
since no computational advantage is being gained. 
In our opinion, investing in a ML model becomes only worthwhile if 
(a) the overall problem is combinatorially hard and costly evaluations are repeatedly required 
while relatively few input variables change by small amounts
(for example, screening solvent mixtures using {\em ab initio} MD calculations);
{\em and } if
(b) the generation of training instances is less costly than solving the entire problem from first principles. 

In this chapter, we review some of the most recent contributions in the field.
More specifically, we first give a brief tutorial summary of the employed ML model in the {\bf Kernel Ridge Regression} section. 
Then we will discuss in section {\bf Representation} 
the various representations (descriptors) used
to encode molecular species, in particular
the molecular Coulomb-matrix (CM), sorted or its eigenvalues, as introduced in~\cite{RuppPRL2012},
and as it can also be used in condensed phase~\cite{MLcrystals_Felix2015},
its randomized permuted sets~\cite{Montavon2013},
a pair-wise bag-of-bonds variant used to study the non-local many-body nature of exponential kernels~\cite{BobPaper},
and a molecular fingerprint descriptor based on a Fourier series of atomic radial distribution functions~\cite{FourierDescriptor}.
We review quantum chemistry data of 134k molecules~\cite{DataPaper2014} in section {\bf Data};
and a recipe for obtaining many properties from a single kernel~\cite{SingleKernel2015} in section {\bf Kernel}.
Subsequently, we discuss ML models of properties of electrons in terms of transmission coefficients in transport~\cite{QuantumTransportML2014} 
and electronic hybridization functions~\cite{LF_DMFT_ML2014,LF_DMFT_ML2015} in section~{\bf Electrons};
as well as chemically accurate ($<$ 1\,kcal/mol) ML models of thermochemical properties, electron correlation energies, reaction energies~\cite{DeltaPaper2015}
and first and second electronic excitation energies~\cite{ML_TDDFTEnrico2015} in section {\bf $\Delta$-Machine Learning}.
In section {\bf Atoms in Molecules} local, linearly scaling ML models for atomic properties such as forces on atoms, 
nuclear magnetic resonance (NMR) shifts, core-electron ionization energies~\cite{MLatoms_2015},
as well as atomic charges, dipole-moments, and quadrupole-moments for force-field predictions~\cite{MTP_Tristan2015}.
ML models of cohesive energies of solids~\cite{MLcrystals_Felix2015,Elpasolite_2015} are reviewed in section~{\bf Crystals}.
Finally, {\bf Conclusions} are drawn at the end.

\section{Kernel Ridge Regression}
\label{sec:method}
\begin{figure}
\centering                                                                                
\includegraphics[width=8.8cm, angle=0.0, scale=1]{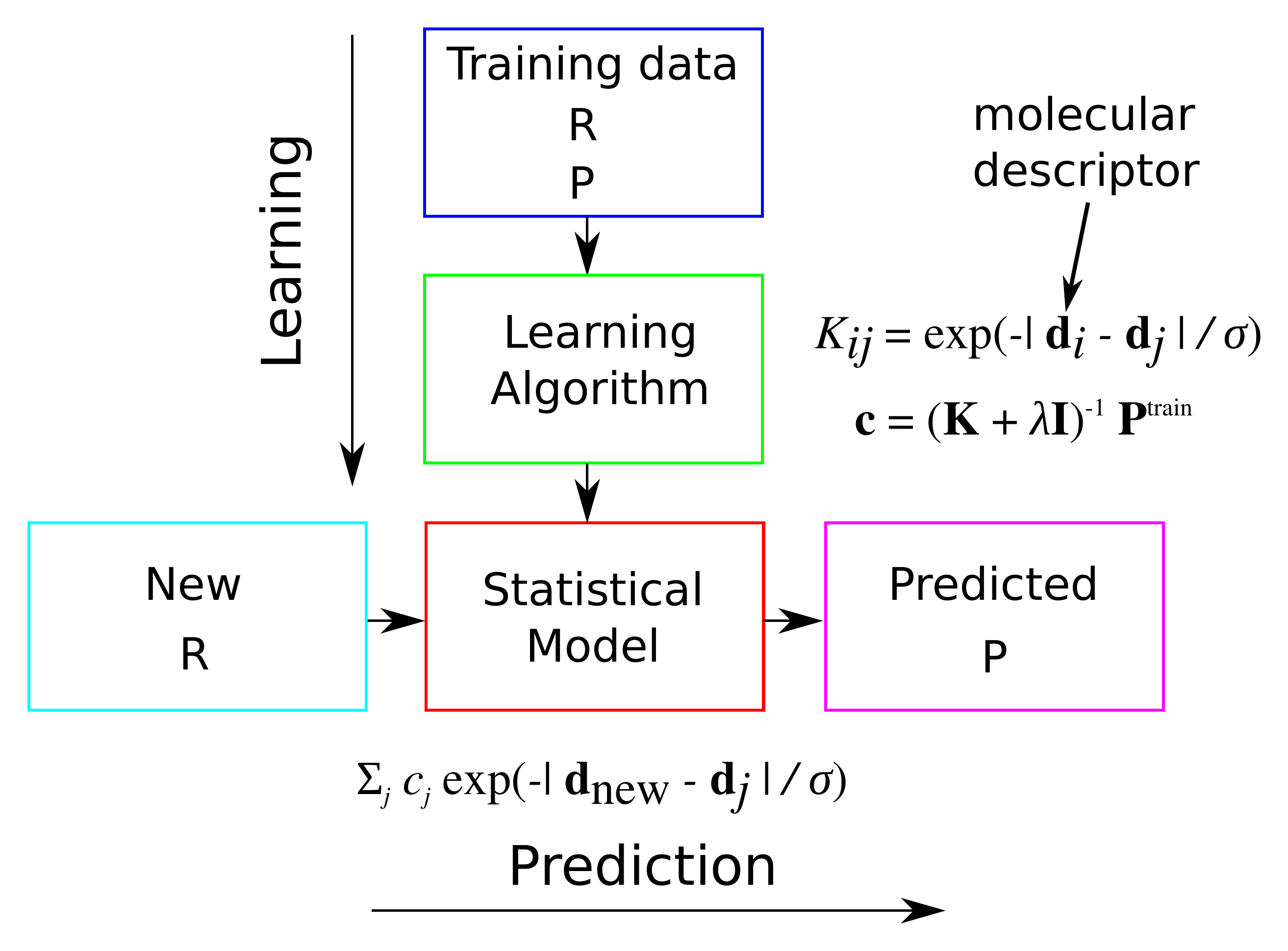}                               
\caption{
Flow chart showing the basic machinery of a Kernel-Ridge-Regression ML setup. 
Vertical flow corresponds to the training the model, calculating regression coefficients
through inversion of the symmetric square kernel matrix with the dimensionality of trainingset size $N$.
Horizontal flow corresponds to out-of-sample predictions of properties of query compounds,
using the regression coefficients to scale the distance dependent contribution of each training instance.
Input compounds are denoted by R, their ML representations, a.k.a.~descriptors, by $\fatd$, and properties by P.
$\sigma$ and $\lambda$ correspond to hyperparameters that control kernel width and noise-level, respectively.
Their optimization is discussed in the {\bf Kernel} section.
 }                                                                                        
\label{fig:flow}                                                                          
\end{figure}

As described previously~\cite{RuppPRL2012,SingleKernel2015,ML_TDDFTEnrico2015},
within the ML approach reviewed herewithin a generic but highly flexible model kernel function
is trained across chemical space with as many free parameters as there are training molecules. 
Other supervised learning models, based on neural networks or random forests, could be used just as well.
We find kernel ridge regression (KRR) appealing because it is simple, robust, and easy to interpret.
In the following, our notations follow Ref.~\cite{SingleKernel2015}. 
We denote matrices by upper bold, and vectors by lower bold cases, respectively.
As long as all variables in SE are accounted for in the molecular representation $\fatd$,
arbitrary numerical accuracy can be achieved by the resulting model of some property $p$---provided that
sufficient and diverse data is made available through training records.
The overall ML setup is schematically presented in Fig.~\ref{fig:flow}. 

This approach 
models some property $p$ of query molecule $q$ as a linear combination of similarities to $N$ training molecules $t$. 
Similarities are quantified as kernel-functions $k$ whose arguments are the
distances $D_{qt}$ between mutual representations $\fatd_q$ and $\fatd_t$ of $q$ and $t$, respectively,
\begin{equation}
p^{\rm target}({\bf d}_q) \; \approx \; \sum_{t=1}^N c_t k(D_{qt})
\label{eq:krr1}
\end{equation}
Using the sorted CM descriptor~\cite{RuppPRL2012} results in accurate ML models of atomization energies 
when combined with a Laplacian kernel function, $k(D_{qt}) = \exp(-D_{qt}/\sigma)$, 
with $D_{qt}$ being the $L_1$ (Manhattan) norm between the two descriptors, $|{\bf d}_q - {\bf d}_t |$, and
$\sigma$ being the kernel width~\cite{AssessmentMLJCTC2013}. 
The regression coefficients, $c_t$, one per training molecule, are obtained through KRR over the training data resulting in
\begin{equation}
({\bf K}+\lambda {\bf I})\ {\bf c} = {\bf p}^{\rm train}.
\label{eq:regression}
\end{equation}
We briefly sketch the derivation of above equation using the least squares error measure 
with Tikhonov regularization~\cite{tikhonov1977solutions,hansen2010discrete}
using a Tikhonov matrix ${\bf K}^m$, where $m$ is usually 0 or 1. 
Let us denote the reference property values of training molecules
as the column vector ${\bf p}^{\rm ref}={\bf x}$. 
The KRR Ansatz for the estimated property values of training molecules is ${\bf p}^{\rm est}={\bf K}{\bf c}$. 
The $L_2$-norm of the residual vector, penalized by regularization of regression coefficients, is the Lagrangian 
\begin{eqnarray}
\mathcal{L}   & = & ||{\bf p}^{\rm ref}-{\bf p}^{\rm est}||_2^2 + \lambda {\bf c}^{\rm T}{\bf K}^m{\bf c} \nonumber \\
             & = & \left( {\bf x}-{\bf K}{\bf c} \right)^{\rm T} \left( {\bf x}-{\bf K}{\bf c} \right) +  
             \lambda {\bf c}^{\rm T}{\bf K}^m{\bf c}   
\end{eqnarray}
 where $(\cdot)^{\rm T}$ denotes transpose operation, and $m$ is a real number.
To minimize the Lagrangian, we equate its derivative with respect to the regression coefficient vector, 
${\bf c}$, to zero
\begin{eqnarray}
\frac{d}{d {\bf c}} \mathcal{L} = 
    -2{\bf x}^{\rm T} {\bf K} + 2 {\bf c}^{\rm T} {\bf K}^2  +   2 \lambda {\bf c}^{\rm T} {\bf K}^m & = & 0.
\label{eq:final}
\end{eqnarray}
Here we have used 
the fact that the kernel matrix ${\bf K}$ is symmetric, i.e., ${\bf K}^{\rm T} = {\bf K}$
along with the identity, 
$\left( d/d{\bf c}   \right){\bf d}^{\rm T}{\bf c}= \left( d/d{\bf c}   \right){\bf c}^{\rm T}{\bf d} = {\bf d}^{\rm T}$, 
where ${\bf c}$ is a column vector
and ${\bf d}^{\rm T}$ is a row vector. After rearranging Eq.~(\ref{eq:final}) and division by 2,  
\begin{eqnarray}
\left(  {\bf K}^2{\bf c} +  \lambda {\bf K}^m{\bf c} - {\bf K}{\bf x} \right)^{\rm T} 
& = &  0. 
\label{eq:final3}
\end{eqnarray}
Taking the transpose and subsequent multiplication from the left with ${\bf K}^{-1}$ results in
\begin{eqnarray}
\left( {\bf K} +  \lambda {\bf K}^{m-1} \right) {\bf c} = {\bf x}.
\label{eq:final5}
\end{eqnarray}
For the values of $m=1$ and 0, we arrive at the equations employed in most KRR based ML-studies employing the 
penalty functions $\lambda {\bf c}^{\rm T}{\bf K}{\bf c}$, and $\lambda {\bf c}^{\rm T}{\bf c}$,
respectively. 
When $m=0$, the penalty rightly favors a regression coefficient vector $\bf c$ with small norm, decreasing
the sensitivity of the model to noisy input. 
On the other hand, this choice of $m$ increases the computational 
overhead by having to compute the ${\bf K}^2$ matrix ---  note that Eq.~(\ref{eq:final3}) for $m=0$ is
\begin{eqnarray}
\left( {\bf K}^2 +  \lambda {\bf I} \right) {\bf c} = {\bf K}{\bf x}.
\end{eqnarray}
For this reason, we typically employ $m=1$ in  Eq.~(\ref{eq:final5}) which leads to the 
regression problem 
\begin{eqnarray}
\left( {\bf K} +  \lambda {\bf I} \right) {\bf c} = {\bf x},
\end{eqnarray}
also stated above in Eq.~(\ref{eq:regression}).

\section{Representation}
\label{sec:Fourier} 
In general, the choice of the representation ${\bf d}$ of the relevant input variables is crucial,
as reflected by the learning rate, i.e.,~by the slope of the out-of-sample error versus trainingset size curve.
The more appropriate the representation, the fewer training instances will be required to reach a
desirable threshold in terms of predictive accuracy. 
In our original contribution~\cite{RuppPRL2012}, we proposed to 
encode molecules through the ``Coulomb''-matrix (CM). 
The genesis of the CM is due to the aim to account for the atomic environment of each atom in a molecule
in a way that is consistent with QM, i.e.,~by using only the information that also enters the molecular Hamiltonian,
and by encoding it in such a way that the molecular Hamiltonian can always be reconstructed. 
A simple way to do so is to use the mutual interactions of each atom $I$ with all other atoms in the molecule. 
Within QM, interatomic energy contributions result from (a) solving the full electronic 
many-body problem (typically within BO approximation and some approximation to the electron correlation) 
and (b) addition of the nuclear Coulomb repulsion energy $Z_I Z_J/|\fatR_I - \fatR_J|$.
As a first step we considered only the latter, and doing this for atom $I$ in the molecule results in 
an array of interaction terms, which becomes a complete atom-by-atom matrix when done for the entire molecule.
While such a $N_I \times N_I$ matrix with unspecified diagonal elements, and containing just Coulomb repulsion terms as
off-diagonal elements, does uniquely represent a molecule, it turns out to be beneficial for the 
learning machine to explicitly encode the atomic identity of each atom 
via the nuclear charge, $Z_I$, or, in order to also have energy units for the diagonal
elements, by the potential energy estimate of the free atom, $E(Z_I) \sim 0.5 Z_I^{2.4}$. 
As noted later~\cite{ML_TDDFTEnrico2015}, these diagonal terms are similar to the
total potential energy of a neutral atom within Thomas-Fermi theory, 
$E_{\rm TF}=-0.77 Z^{7/3}$, or its
modifications with a $Z$-dependent prefactor in the range 0.4--0.7~\cite{parryang}. 

Note that the CM is a variant of the adjacency matrix, 
i.e.~it contains the complete atom-by-atom graph of the molecular geometry with off-diagonal elements
scaling with the inverse interatomic distances.
Inverse interatomic distance matrices have already previously been proposed as descriptors of molecular
shape~\cite{TodeschiniConsonniHandbookDescriptor}.
As the molecules become larger, or when using off-diagonal elements with higher orders of the inverse distance , 
such representations turn into sparse band matrices. 
Periodic adaptations of the CM have also been shown to work for 
ML models of condensed phase properties, such as cohesive energies~\cite{MLcrystals_Felix2015}.
Generation of the CM only requires interatomic distances, 
implying invariance with respect to translation and rotational degrees of freedom. 
The CM is also differentiable with respect to nuclear charge or atomic position,
and symmetric atoms have rows and columns with the same elements.
Consequently, while the full spatial and compositional information of any molecule is being accounted for,
the global chirality is lost: Left- and right-handed amino acids have the same CM.
Diastereomers, however, can be distinguished.
The indexing of the atoms in the CM is arbitrary, hence
one can introduce invariance with respect to indexing by using the $N_I$ eigenvalues of the CM, as it was originally done in Ref.~\cite{RuppPRL2012}.  
The eigenvalues, however, do not uniquely represent a system, as illustrated in Ref.~\cite{MoussaComment}.
Albeit never encountered in the molecular chemical space libraries we employed so far, this issue can 
easily be resolved by using a CM with atom-index sorted by the $L_2$ and $L_1$ norms of each row:
First, we sort the ordering based on $L_2$  norm~\cite{MoussaReply}. 
Second, whenever $L_2$ norms of two rows are degenerate by coincidence, 
they can be ordered subsequently based on their $L_1$ norms. 
This choice is justified by the fact that only symmetric atoms will be degenerate for both norms.
While the CM itself is differentiable, this sorting can lead to distances between CMs which are not differentiable. 
The Frobenius norm between two CM matrices, for example, 
will no longer be differentiable whenever two non-symmetric (differing $L_1$ norms of CM rows)
atoms in one CM reach degeneracy in the $L_2$ norm of their rows. 
In other words, for this to happen, two non-symmetric atoms in the same molecule would have to have the same (or very similar)
atomic radial distribution functions (including all combinations of atom-types).  
We have not yet encountered such a case in any of the molecular chemical space libraries we employed so far.
However, while molecules corresponding to compositional and constitutional isomers appear to be sufficiently 
separated in chemical space this issue of differentiability might well limit the applicability of 
Frobenius based similarities that rely on CM representations, or any other adjacency matrix based descriptor for that matter. 
Dealing with conformational isomers, or reaction barriers, could therefore require additional constraints
in order to meet all the criteria of the aforementioned paradigm. 

A simple brute-force remedy consists of using sets of all Coulomb matrices one can obtain by
permuting atoms of the same nuclear charge. Unfortunately, for larger molecules this solution becomes
rapidly prohibitive.  
For medium sized molecules, however, randomized permuted subsets have yielded some good results~\cite{Montavon2013}.

Recently, based on the external potential's Fourier-transform, a new fingerprint descriptor has been proposed 
which meets all the mathematically desirable requirements: It uniquely encodes molecules 
while conserving differentiability as well as invariance with respect to rotation, translation, and atom indexing~\cite{FourierDescriptor}.
This descriptor corresponds to a Fourier series of atomic radial distribution functions $RDF_I$ on atom $I$, 
\beq
{\bf d} \propto \sum_I Z_I^2 \cos[RDF_I],
\eeq
where linearly independent atomic contributions are obtained (non-vanishing Wronskian). 
Alas, while this representation does result in decent ML models, its predictive accuracy is inferior to 
the atom-sorted CM for the datasets of organic molecules tested so far.
A more recently proposed alternative scatter transform based descriptor also succeeds in uniting all the desirable formal 
properties of a good descriptor, but so far it has only been shown to work for planar molecules~\cite{Mallat_wavelet2014}.

If we are willing to compromise on the uniqueness property, a pair-wise variant of the CM, called
bag-of-bonds (BOB) corresponding simply to the set of vectors with all off-diagonal elements for all
atom-pair combinations, can be constructed~\cite{BobPaper}. 
Unfortunately, non-unique descriptors can lead to absurd results, as demonstrated in Ref.~\cite{FourierDescriptor}.
In the case of BOB, arbitrarily many geometries can be found (any homometric pairs) for which a BOB kernel will yield
degenerate property predictions. 
Since homometric molecules differ in energy by all interatomic many-body terms, this can lead to errors
substantially larger than chemical accuracy.
But also the predictive accuracy of ML models for other properties than energies will suffer from representations which cannot distinguish homometric molecules. 
For example, consider the following homometric case of two clusters consisting of four identical atoms. 
The geometry of one cluster corresponds to an equilateral triangle with edge-length A and with the fourth atom in the center at interatomic distance B. 
The geometry of the second cluster corresponds to a distorted tetrahedron with one side being spanned by an equilateral triangle with edge-length B,
and the fourth atom being located above the triangle at interatomic distance A to the other three atoms.
These two clusters will have very different multi-pole moments which will be degenerate by construction within
{\em any} ML model that relies on lists of pair-wise atom-atom distances without any additional connectivity information.
It is important to note, therefore, that non-unique representations can lead to ML models which do {\em not}
improve through the addition of more data---defying the original purpose of applying statistical learning to chemistry. 
Nevertheless, in the case of organic molecules drawn from GDB~\cite{GDB17}, BOB has shown slightly superior performance 
for atomization energies, frontier orbital eigenvalues as well as polarizabilities. 
The fact that it can even effectively account for many-body energy 
contributions through non-linear kernels, thereby outperforming any explicit effective pair-wise potential model, 
has also been highlighted~\cite{BobPaper}. 
For extended or periodic systems, however, we expect the likelihood of encountering homometric (or near-homometric)
configurations to increase, and therefore the implications of lacking uniqueness to become more important.

\section{Data}
\begin{figure*}
\centering
\includegraphics[width=12.5cm, angle=0.0, scale=1]{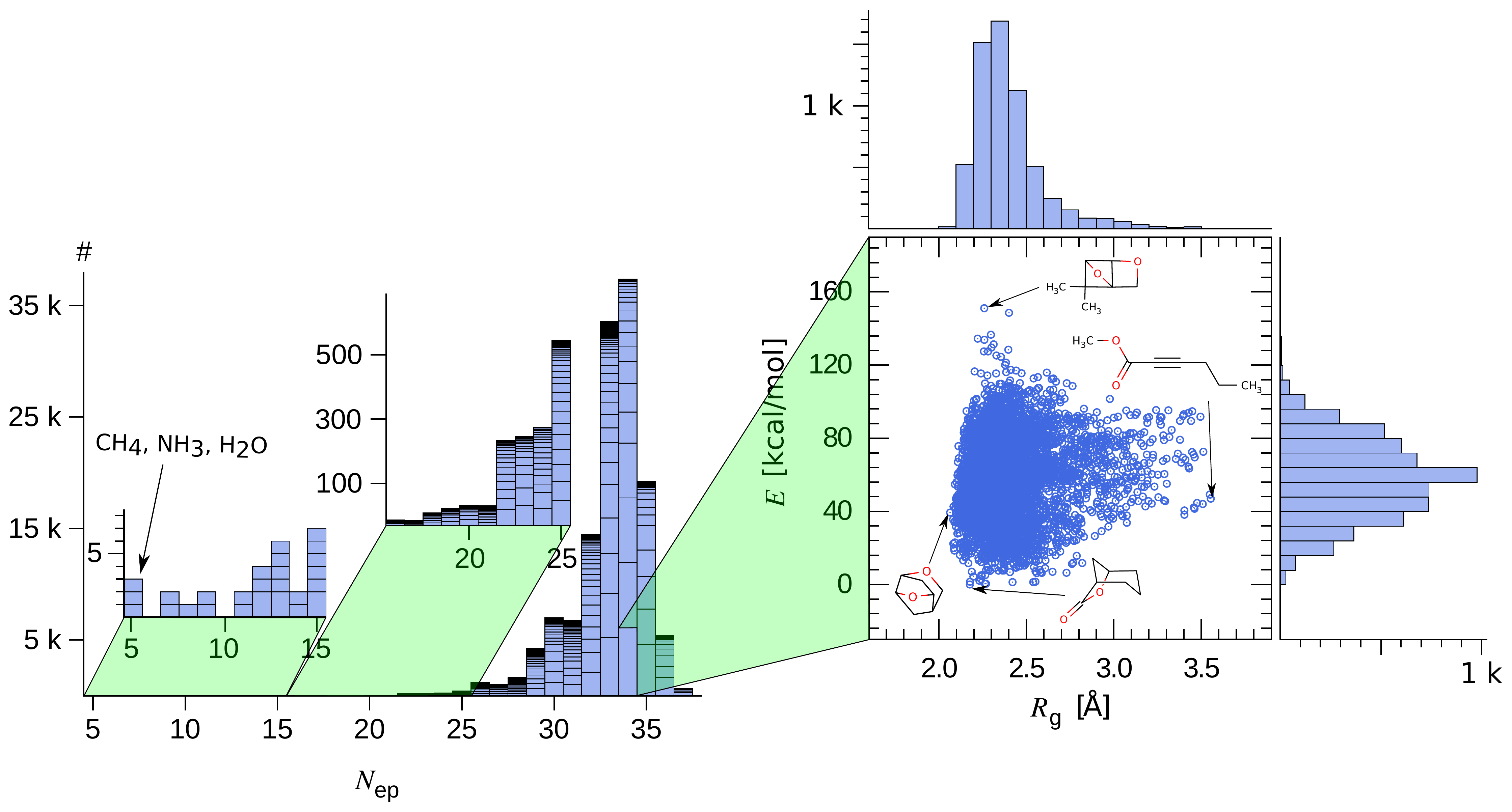}
\caption{
Distribution of molecular size (in terms of number of occupied orbitals, i.e.~electron pairs, $N_{\rm ep}$).
The height of each black box denotes the number of constitutional isomers for one out of 621 stoichiometries present in 134\,k molecules. 
The two left-hand side insets correspond to zoom-ins for smaller compounds. 
The right-hand side inset zooms in on the predominant stoichiometry, C$_7$H$_{10}$O$_2$, and
features a scatter plot of G4MP2 relative (w.r.t. global minimum) potential energies of atomization $E$ versus
molecular radius of gyration, $R_g$, as well as joint projected distributions. 
(Reprinted from Ref.~\cite{DataPaper2014}. Copyright (2014), Nature Publishing Group.)
}\label{fig:134kstoi}
\end{figure*}

With the growing interest in developing new and efficient ML strategies, all aiming towards improving 
the predictive power to eventually reach standard quantum chemistry accuracy and transferability at negligible computational cost, 
it has become inevitable to employ large molecular training sets. 
To provide a standard benchmark dataset for such quantitative comparative studies, we  
have produced one of the largest quantum chemistry libraries available, dubbed QM9,
containing structures and properties obtained from quantum chemistry calculations for 133,885 (134\,k)
small organic molecules, containing up to 9 CONF atoms~\cite{DataPaper2014}. 
The 134\,k QM9 set is a subset of the GDB-17 database published by Reymond {\it et al}.~\cite{GDB17}, 
which contains a list of molecular graphs of 166,443,860,262 (166\,G) synthetically feasible molecules. 
The chemical diversity of QM9 is illustrated in Fig.~\ref{fig:134kstoi}. 
The possibility to generate quantum chemistry Big Data such as the 134\,k dataset has
also stimulated efforts towards designing efficient formats for the archival and retrieval of 
large molecular datasets \cite{alvarez2014managing,harvey2015standards}. 

The starting point for generating the QM9 data published in Ref.~\cite{DataPaper2014} 
was the original set of SMILES descriptors of the smallest 134\,k molecules in GDB-17. 
Initial Cartesian coordinates were obtained using the program CORINA~\cite{CORINA}. 
The semi-empirical quantum chemistry method PM7~\cite{PM7}, and density functional theory~\cite{KS} 
using the B3LYP~\cite{B3LYP} exchange-correlation potential with the basis set 6-31G(2df,p), 
were subsequently employed to relax the CORINA structures into local energy minima, 
using software programs MOPAC~\cite{MOPAC} 
and Gaussian09 (G09)~\cite{ftssrcsbmpetal2009}, respectively.  
For all molecules, this dataset contains multiple properties, computed at the B3LYP/6-31G(2df,p) level, including principal moments of inertia, dipole moment, 
isotropic polarizability, energies of frontier molecular orbitals, radial expectation value, harmonic vibrational wavenumbers, 
and thermochemistry quantities heat capacity (at 298.15~K), internal energy (at 0~K and 298.15~K), enthalpy (at 298.15~K), and free energy (at 298.15~K). 
The dataset also includes reference energetics for the atoms H, C, N, O, and F to enable computation of chemically more relevant energetics such as 
atomization energy, and atomization heat (enthalpy). 
For 6095 (6\,k) constitutional isomers with the stoichiometry C$_7$H$_{10}$O$_2$, thermochemical properties
have also been computed at the more accurate G4MP2 level~\cite{G4MP2}.  
B3LYP atomization heats deviate from G4MP2 with a mean absolute deviation of $\sim$5.0 kcal/mol. 

For recent ML studies~\cite{rampiFP,BobPaper,DeltaPaper2015,FourierDescriptor,MLatoms_2015,SingleKernel2015,ML_TDDFTEnrico2015,PavloML}, the QM9 134\,k dataset 
has served as a standard benchmark providing diverse organic molecules and properties. 
It has been crucial in order to convincingly demonstrate the transferability of ML models' performance to thousands of out-of-sample molecules, that were not part of training. 
While these ML studies have used the structures, and properties from QM9 without modifications, 
it has sometimes been necessary to extend the dataset by including additional quantities/structures. 
For the $\Delta$-ML study discussed below
9868 (10\,k) additional diastereomers for the stoichiometry C$_7$H$_{10}$O$_2$ have been generated for 
additional transferability tests~\cite{DeltaPaper2015}. 
For the 22\,k molecules constituting the QM8 subset (all molecules with exactly 8 CONF atoms taken out of the QM9 134\,k set) 
low-lying electronic spectra (first two singlet-singlet excitation energies, and oscillator strengths) 
computed at the levels, linear-response time-dependent DFT (LR-TDDFT) \cite{LRTDDFT}, and linear response approximate second-order 
coupled cluster method (CC2) \cite{CC2} have been computed using the code TURBOMOLE.~\cite{ML_TDDFTEnrico2015} 
At both levels, Ref.~\cite{ML_TDDFTEnrico2015} presented results with the large basis set def2TZVP \cite{def2basis}, and DFT results included
predictions based on exchange-correlation potentials PBE~\cite{PBE}, PBE0~\cite{PBE0,PBE0Barone}, and CAM-B3LYP~\cite{CAMB3LYP}. 
CC2 calculations employed the resolution-of-the-identity (RI) approximation for two-electron repulsion integrals.

\section{Kernel}
Conventionally, one of the more repetitive aspects of KRR based studies for small training
sets consists of the optimization of hyperparameters (kernel width, $\sigma$, and regularization strength (or noise-level), $\lambda$).
This is typically accomplished following time-consuming, 5 or 10-fold, cross-validation (CV) procedures~\cite{RuppTutorial2015}. 
Albeit easily parallelizable, it is a task which can also conveniently be circumvented using simple heuristics:
Within various studies training ML models on quantum chemistry results
it has been noticed that CV procedures result in very small, if not negligible, $\lambda$ values. 
Setting $\lambda$ directly to zero typically does not yield deteriorated learning rates. 
Since $\lambda$, also known as "noise-level", is supposed to account for noise in the training data 
we do not consider this observation to be surprising: 
Observables calculated using numerical solutions to given approximations to Schr\"odinger's equation do not suffer from any noise 
other than the numerical precision---multiple orders of magnitude smaller than the properties of interest. 

It is also straightforward to choose the exponential kernel width: 
A decent choice is the median of all descriptor distances in the training set~\cite{MLbook}, 
i.e., $\sigma=cD_{ij}^{\rm median}$, where  $c=\mathcal{O}(1)$. 
Snyder {\it et al.} \cite{snyder2013orbital} used this choice for modeling density functionals with a Gaussian kernel. 
This choice of $\sigma$ can be problematic when descriptor distances exhibit a non-unimodal density distribution. 
In Ref.~\cite{SingleKernel2015}, we interpreted the role of $\sigma$ as a coordinate scaling 
factor to render ${\bf K}$ well-conditioned. 
This suggestion is based on the fact that for  $\sigma \approx 0$, off-diagonal elements of 
the kernel matrix, $K_{ij}^{\rm Laplace}=\exp\left(-D_{ij}/\sigma\right)$
or $K_{ij}^{\rm Gauss}=\exp\left(-D_{ij}^2/(2 \sigma^2)\right)$
vanish, resulting in a unit-kernel-matrix, ${\bf K}={\bf I}$. 
On the other hand, for $\sigma >> 1$, a kernel-matrix of ones is obtained which would be singular. 
Instead, one can select a $\sigma$ value such that the lower bound of the kernel elements is 0.5~\cite{SingleKernel2015}.
For Laplacian and Gaussian kernels, this implies constraining the smallest kernel matrix element, 
which corresponds to the two most distant training molecules, to be 1/2, i.e.,
${\rm min}(K_{ij}) = f( {\rm max}(D_{ij}), \sigma) = 1/2$, and solving for $\sigma$.
For the Laplacian kernel, 
$\sigma_{\rm opt}^{\rm Laplace} = {\rm max}(D_{ij}) / \log(2)$, and for the
Gaussian kernel, 
$\sigma_{\rm opt}^{\rm Gauss} = {\rm max}(D_{ij}) / \sqrt{2\log(2)}$. 
For the B3LYP/6-31G(2df,p) molecular structures in the 134\,k set with CHONF stoichiometries, 
${\rm max}(D_{ij})=1189.0 $ a.u., 
and 
${\rm max}(D_{ij})=250.4 $ a.u.,
when using $L_1$ (Manhattan) and $L_2$ (Euclidean) distance metrics, for sorted CM descriptors, respectively. 
As one would expect for this data set and this descriptor, maximal distances are found for the molecular pair methane and hexafluoropropane.
These distances correspond to $\sigma_{\rm opt}^{\rm Laplace} = 1715$ a.u. (with $L_1$ distance metric), and 
$\sigma_{\rm opt}^{\rm Gauss} = 213$ a.u. (with $L_2$ distance metric). 
These values are of the same order of magnitude as for a molecular dataset similar to the 134\,k set,
i.e.~$\sigma_{\rm opt}^{\rm Laplace} \thickapprox 1000$ a.u. employed in \cite{SingleKernel2015}, 
and $50 < \sigma_{\rm opt}^{\rm Gauss} < 500$ a.u., as also noted in \cite{anatole-ijqc2013}. 

The numerical robustness of choosing $\sigma$ exclusively from the distance distribution can be rationalized using precision arguments. 
When the lower bound of elements in ${\bf K}$ is limited to $q$, i.e., $q \le K_{ij} \le 1$, the lower bound of the  
conditional number $\kappa$ of ${\bf K}$ follows $ \kappa \ge 1 + N q/(1-q) $, 
where $N$ is the trainingset size (number of rows/columns of ${\bf K}$). 
This condition is valid for any dataset. 
$\kappa$ is the ratio between the maximal and minimal eigenvalues of ${\bf K}$.
Typically, when $\kappa= 10^t$, one looses $t$ digits of accuracy in the ML model's prediction, 
even when employing exact direct solvers for regression. 
When using the single-kernel {\it Ansatz}~\cite{SingleKernel2015} 
with $q=1/2$,  $\kappa$ can be as small as $1+N$, i.e., one can expect at best a numerical precision of $\sim\log[N+1]$ digits.
Depending on the desired precision, consideration of $\kappa$ to dial in the appropriate $q$ can be important. 

Using hyperparameters which are derived from a representation which uses exclusively chemical composition 
and structure rather than other properties, has multiple advantages. 
Most importantly, the global kernel matrix becomes property-independent! 
In other words, kernel matrices inverted once, ${\bf K}^{-1}$, can be used to generate regression coefficients 
for arbitrarily varying properties e.g.~functions of external fields~\cite{QuantumTransportML2014,LF_DMFT_ML2014,LF_DMFT_ML2015}.
Furthermore, ML models for arbitrarily different molecular properties, resulting from expectation values of different 
operators using the same wavefunction, can be generated in parallel using the same inverted kernel. 
More specifically,
\begin{eqnarray}
\left[ {\bf c}^{p_1} {\bf c}^{p_2} \ldots {\bf c}^{p_n} \right]= 
{\bf K}^{-1} \left[ {\bf p}_1 {\bf p}_2 \ldots {\bf p}_n \right],
\label{eq:inv2}
\end{eqnarray}
where ${\bf p}_1, {\bf p}_2, ..., {\bf p}_n$, are $n$ property vectors of training molecules.
The inverse of the property independent, single-kernel once obtained can also be stored 
as described in  Ref.~\cite{SingleKernel2015} to enable reuse.
Alternatively, the kernel matrix can be once factorized as lower and upper triangles by, say,
Cholesky decomposition, ${\bf K}={\bf L}{\bf U}$, and the factors can be stored. The regression coefficient
vectors can then be calculated through a forward substitution followed by a backward substitution 
\begin{eqnarray}
{\bf L}  \left[ {\bf y}^{p_1} {\bf y}^{p_2} \ldots {\bf y}^{p_n} \right] & = & 
\left[ {\bf p}_1 {\bf p}_2 \ldots {\bf p}_n \right] \nonumber \\
{\bf U}  \left[ {\bf c}^{p_1} {\bf c}^{p_2} \ldots {\bf c}^{p_n} \right] & = & 
\left[ {\bf y}^{p_1} {\bf y}^{p_2} \ldots {\bf y}^{p_n} \right].
\label{eq:forward}
\end{eqnarray}

\begin{figure*}
\centering
\includegraphics[width=12.5cm, angle=0.0, scale=1]{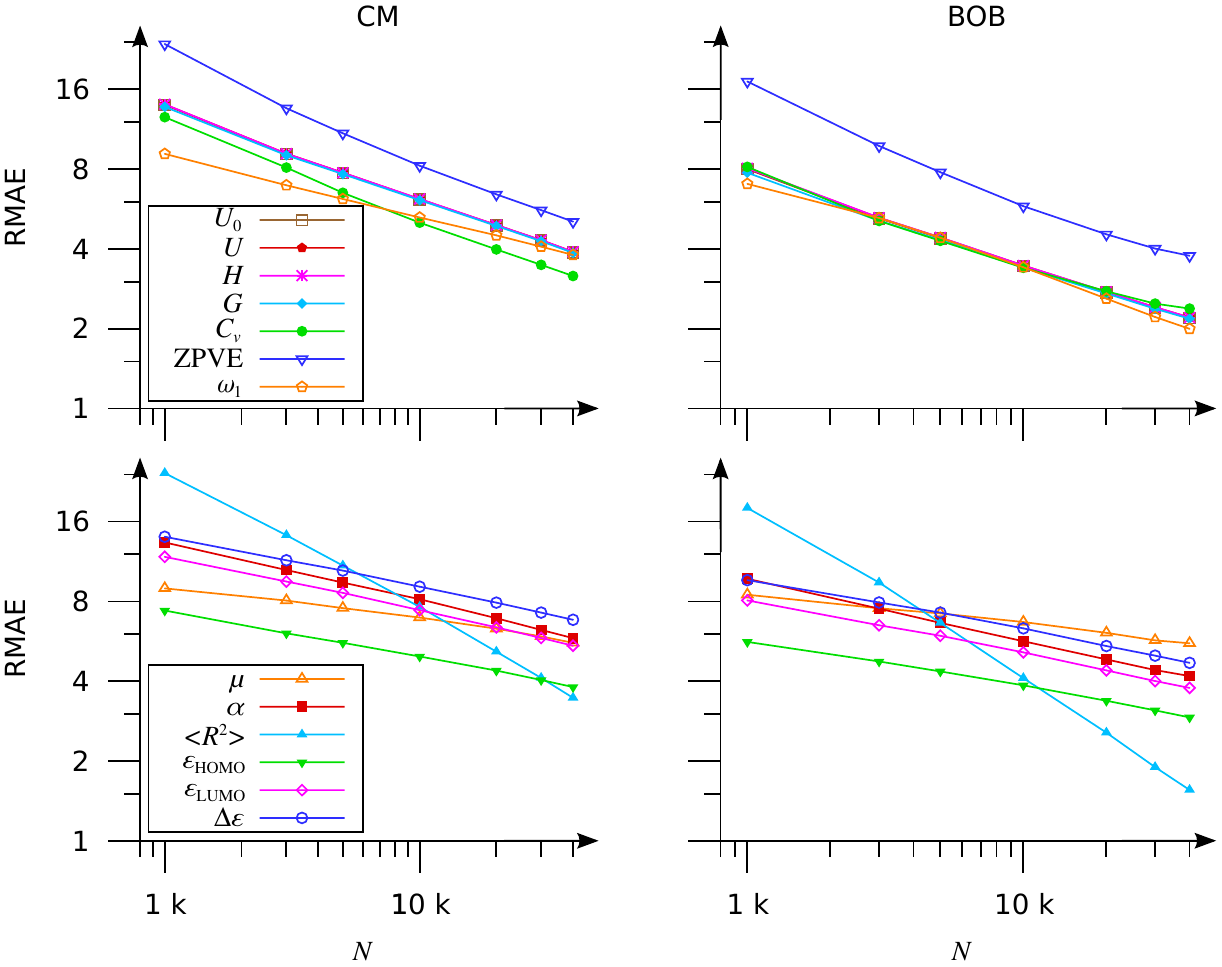}                               
\caption{
Learning curves for single-kernel ML models (with $\sigma=1000$ a.u. and $\lambda=0$) for
thirteen molecular properties including thermochemical (TOP) as well as electronic (BOTTOM) properties of molecules. 
For each trainingset size, a single kernel has been inverted and trained and tested on thirteen properties.
Left and Right refer to the use of the aforementioned descriptors Coulomb matrix (CM) and bag-of-bonds (BOB), respectively.
Out-of-sample relative mean absolute errors (RMAE), i.e., MAE relative to standard quantum chemistry accuracy thresholds, 
 are plotted for test and training sets drawn from the 112\,k GDB-17 subset of molecules with exactly 9 atoms CONF~\cite{GDB17}.
See Ref.~\cite{SingleKernel2015} for further details of quantum chemistry accuracy thresholds. 
See Refs.~\cite{SingleKernel2015,BobPaper} for a color version.
} \label{fig:cm_vs_bob}
\end{figure*}

The single kernel idea has been used (with $\sigma=1000$ a.u.~and $\lambda=0$),
to generate 13 different molecular property models for all of which
out-of-sample prediction errors have been shown to decay systematically with trainingset size~\cite{SingleKernel2015}.
These learning curves are on display in Fig.~\ref{fig:cm_vs_bob}, and have been obtained using the QM9 data set
of 134\,k organic molecules for training and testing~\cite{DataPaper2014}. 
In order to compare errors across properties, relative mean absolute errors (RMAE) 
with respect to desirable quantum chemistry accuracy thresholds, have been reported. 
For instance, the target accuracy for all energetic properties is 1 kcal/mol; 
thresholds for other properties, as well as further technical details, can be found in Ref.~\cite{SingleKernel2015}. 

Note the systematic decay in RMAEs for {\it all} molecular properties in Fig.~\ref{fig:cm_vs_bob}, independent of choice of descriptor. 
We consider this numerical evidence that accurate ML-models for arbitrarily many QM properties can be built
using an exclusively structure based single kernel matrix---as long as the latter spans a chemical
space which is representative of the query compound. 
While none of the properties were predicted  within the desirable RMAE of 1, not even for 
the largest trainingset size $N$=40\,k, we note that the linear nature of the results
suggests that such accuracy thresholds can be achieved by simply increasing the training set size. 

Another advantage of the single kernel {\em Ansatz} is that it enables numerical comparison
of ML models of different properties, as well as different representations.
Most interestingly, some properties are easier to learn than others. 
The electronic spread $\langle R^2 \rangle$ exhibits a noticeably steeper learning rate than
all other properties, starting off with the worst out-of-sample predictions for 1k-ML models, 
and yielding best performance for the 40k-ML models.
The magnitude of the molecular dipole moment, $\mu$, was found to exhibit the smallest learning
rate, with the CM based kernel yielding slightly better results than the BOB descriptor. 
For all other properties, however, the BOB descriptor yields a slightly better performance.
However, as already pointed out above, BOB only accounts for all pair-wise distances and therefore
suffers from lack of uniqueness, and any non-unique representation can lead to absurd ML results~\cite{FourierDescriptor}.
This shortcoming appears to show up for the BOB based 40k-ML models of energetic properties: 
The learning rate levels off and it is not clear if any further improvement in accuracy can be obtained by merely adding more data. 

Conceptionally, making the distinction of properties corresponding to regression coefficients, 
and chemical space to the kernel is akin to the well established distinctions made within the postulates of
quantum mechanics. Namely that between Hamiltonian encoding structure and composition (and implicitly also wavefunction) 
and some observable's operator (not necessarily Hermitian, resulting from correspondence principle---or not). 
The former leads to the wavefunction via Ritz' variational principles, 
which is also the eigenfunction of the operators enabling the calculation of property observables as eigenvalues. 
As such, Ref.~\cite{SingleKernel2015} underscores the possibility to establish an
inductive alternative to quantum mechanics, 
given that sufficient observable/compound pairs have been provided. 
One advantage of this alternative is that it sidesteps the need for many of the explicit manifestations of
the postulates of quantum mechanics, such as Schr\"odinger's equation.
The disadvantage is that in order to obtain models of properties, experimental training values would
have to be provided for all molecules represented in the kernel.

\section{Electrons}
\label{sec:Transmission}
In Ref.~\cite{QuantumTransportML2014} ML models of electron transport properties have been introduced. 
More specifically, transmission coefficients are being modeled for a hexagonal conduction channel consisting of atoms with one orbital. 
Such systems are representative for carbon nano-ribbons, and might have potential applications as molecular wires. 
For variable kinetic energy incoming electrons are being injected through
two ballistic leads, coupled on each side of the wire without reflection.
The wire corresponds in length to four times an armchair graphene nanoribbon unit cell, and in width to fourteen atoms.
Depending on composition and geometry of five randomly distributed impurities in the nano-ribbon, 
the transmission probability of electrons (being either transmitted or scattered back) can vary. 
Reference values are obtained for multiple energy values and thousands of compounds 
using a tight-binding Hamiltonian, the Landauer-B\"uttiker approach, and advanced/retarded Green's functions~\cite{INSPEC:1957A08595,INSPEC:143734,PhysRevLett.57.1761}
to calculate the transmission probability~\cite{PhysRevB.23.6851}.
ML models have been trained for three experiments, 
(i) at fixed geometry, the impurities only differ through their coupling (hopping rates) to adjacent 
sites (off-diagonal elements $\gamma_{ij}$ in the Hamiltonian),
(ii) at fixed geometry, impurity elements in the Hamiltonian also differ in on-site energies ($\epsilon_i$).
(iii) as (ii) but with randomized positions.
The nano-ribbon impurity representation $\fatd$ is derived from the Hamiltonian, and inspired by the 
CM: It is a matrix containing the on-site energies along the diagonal, and
hopping elements or inverse distances as off-diagonal elements. 
It is sorted by the rows' norm in order to introduce index invariance (enabling the removal of symmetrically redundant systems).
The kernel function is a Laplacian, and its width is set to a constant, characteristic to the length-scale of training instance distances. 
The kernel width (noise-level) is set to zero.
For all three experiments, training sets with 10,000 (10\,k) randomized examples have been calculated. 
In each case the prediction error of ML models of transmission coefficients is shown to decrease
systematically for out-of-sample impurity combinations as the trainingset size is being increased to up to $N=$8\,k instances. 
Furthermore, as one would expect, the prediction error increases with the complexity of the function
that is being modeled, i.e.,~in experiment (i) it is smaller than in (ii), for which it is smaller than in (iii). 

\begin{figure*}
\centering
\includegraphics[width=12.5cm, angle=0.0, scale=1]{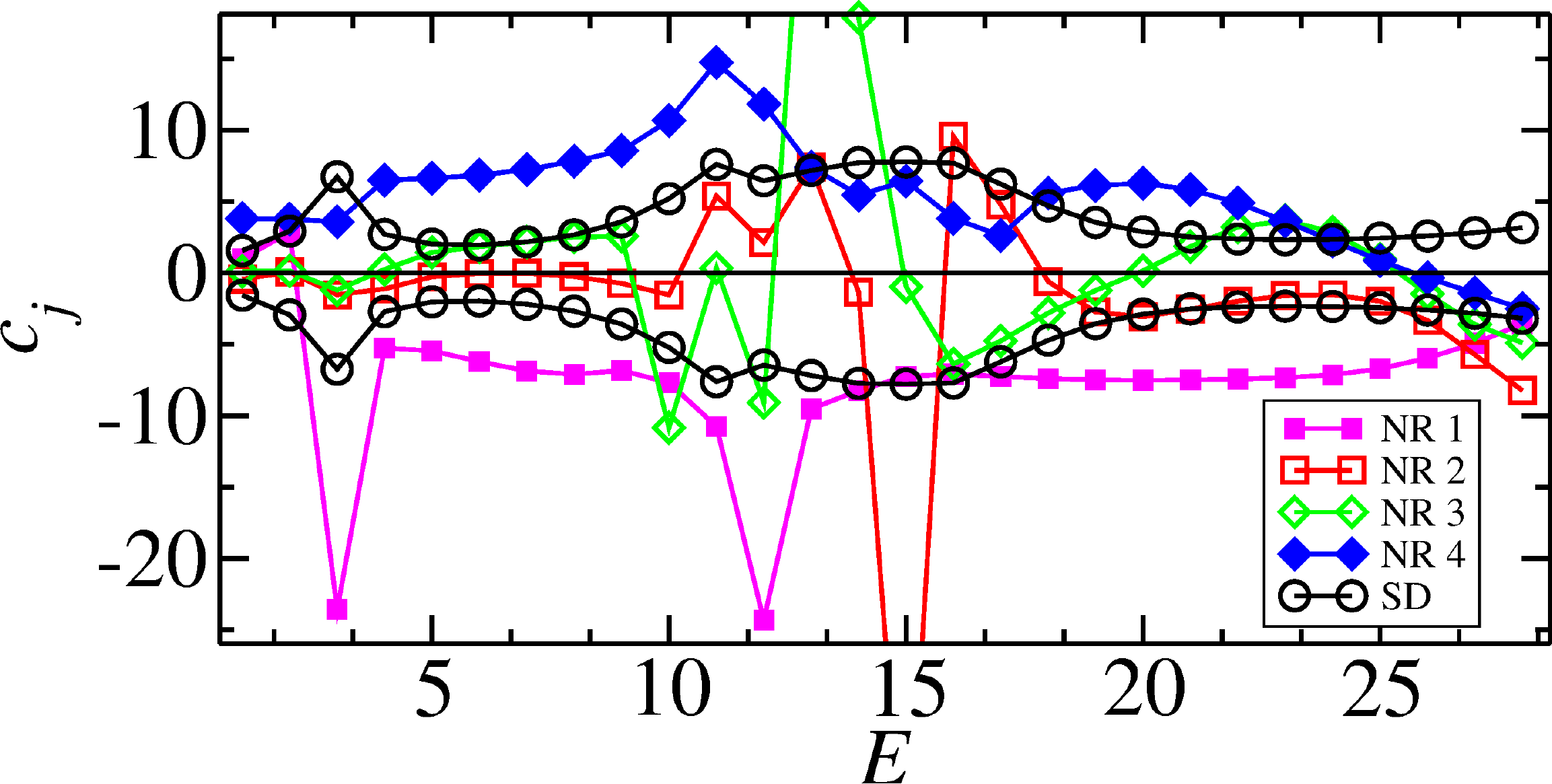}
\caption{
Training coefficients in ML models of transmission coefficients plotted as a function of kinetic energy index 
of the incoming electron, as published in Ref.~\cite{QuantumTransportML2014}.
Shown are the coefficients of four (out of 8000) nano-ribbon compounds used for training (NR 1-4). 
The materials were selected because their coefficient exhibit strongest (NR 2 and NR 3) or weakest (NR 1 and 4) fluctuations in $E$.
The standard deviation (SD) over all 8000 training compounds is shown as well.
}\label{fig:alfas}
\end{figure*}

Another interesting aspect of said study is the modeling of $T$ not only as a function of impurities, 
represented in $\fatd$, but also as a function of an external parameter, namely the kinetic energy of the incoming electron, $E$. 
For each of the three experiments, $T$ was discretized on an $E$ grid, and ML models have been obtained for each $E$ value. 
Since the kernel is independent of $E$, this implies that the role of learning a property as a function 
of some external parameter is assumed by the regression coefficients, while the role of varying impurities is assumed by the kernel.
More specifically, the effective model was
$T^{\rm est}(\fatd_{\rm new},E) = \sum_j c_j(E) \exp(-|\fatd_{\rm new}-\fatd_j|/\sigma)$.
The resulting coefficients $\{c_j(E)\}$ are functions of $E$ for any of the training compounds. 
Exemplary $\{c_j(E)\}$ are on display in Fig.~\ref{fig:alfas} for the two nano-ribbon training compounds for which
coefficients vary the most, as well as for the two nano-ribbons for which they vary the least.
Overall, the variation of the coefficients is rather smooth, suggesting a ``well-tempered''
ML model within which the role of each training instance responds reasonably to external tuning of the target function.
However, when it comes to energy values in the vicinity of the step-wise variation of the transmission coefficient,
(energy indices 12 to 15), strong fluctuations are observed for all four parameters. 
We note that also the standard deviation (corresponding to the distributions shown as insets in Fig.~(2) in Ref.~\cite{QuantumTransportML2014}),
coincides with this movement. 
In Refs.~\cite{LF_DMFT_ML2014,LF_DMFT_ML2015} similar ideas have been applied to 
the modeling of hybridization functions, relevant to dynamical mean field theory.

\section{$\Delta$-Machine Learning}
\begin{figure*}
\centering
\includegraphics[width=12.5cm, angle=0.0, scale=1]{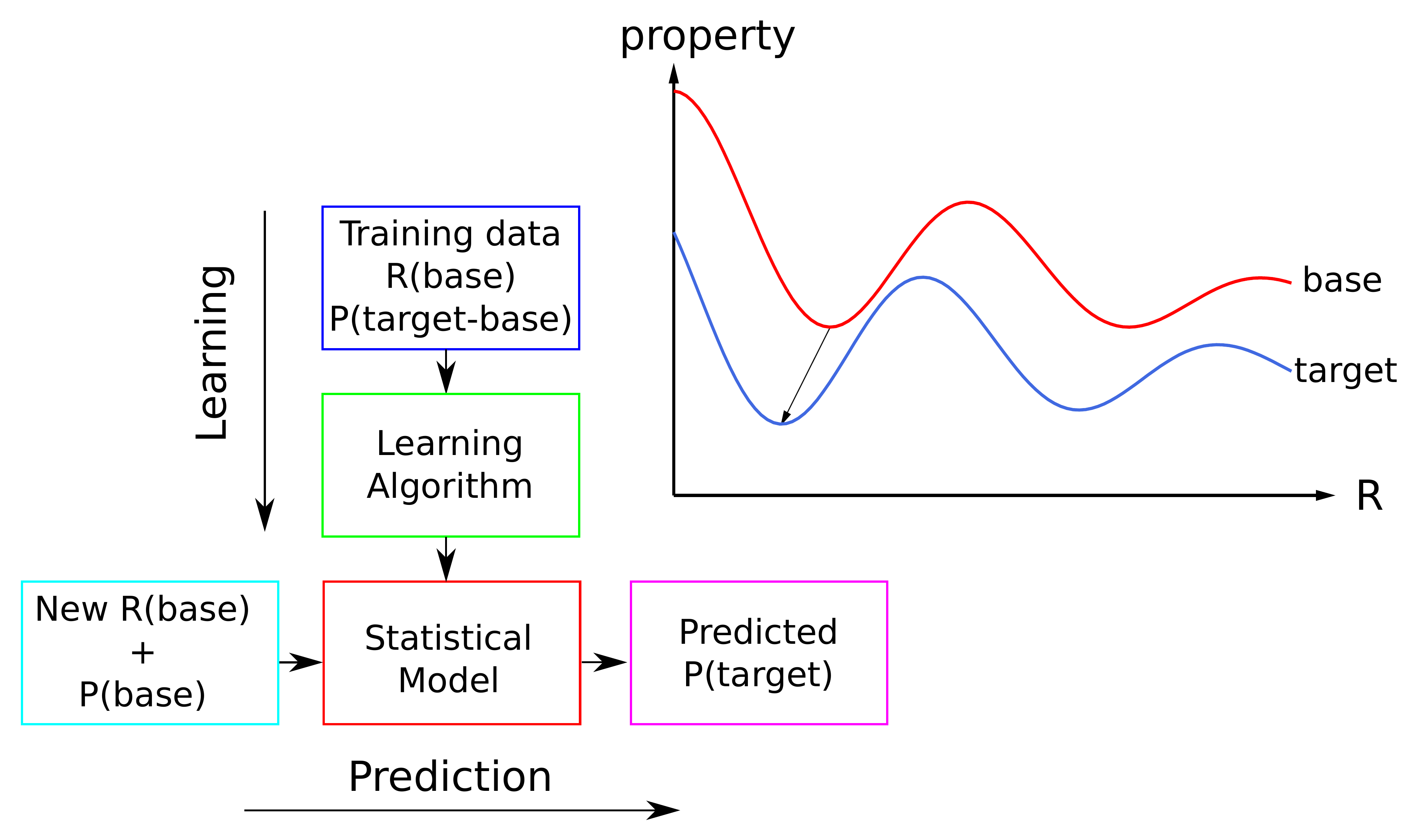}                               
\caption{
Left: Flow chart for $\Delta$-machine learning. 
The input to the model typically consists of structures at a baseline level, $R({\rm base})$, 
and properties are modeled using training data corresponding to properties {\em and} geometries consistent with targetline level, 
$R({\rm target})$.
Right: Baseline and targetline property surfaces, the $\Delta$-ML model (arrow) accounts for both, 
difference in property and geometry due to difference in level of theory.
}
\label{fig:flowdelta}
\end{figure*}

An efficient strategy to improve the predictive accuracy of ML models for computational chemistry
applications, is to augment predictions from inexpensive baseline theories with ML models that 
are trained on the error in the baseline model's prediction with respect to a more accurate targetline prediction. 
This approach, introduced as ``$\Delta$-ML'' \cite{DeltaPaper2015}, 
assumes that baseline and targetline predictions have good rank-ordering correlation. 
A flow chart illustrating the idea is featured in Fig.~\ref{fig:flowdelta}.
The choice of baseline model can tremendously benefit from expert knowledge, 
such as DFT approximations following the hierarchy of ``Jacob's ladder'' \cite{jacobsladder},
or that the accuracy of correlated many-body methods typically increases with the scaling of computational cost, e.g. 
CCSD being more accurate/expensive than MP2 than HF. 
But also the first order approximation to the lowest excitation energy 
using just the HOMO-LUMO gap from single reference theories has shown promise.
However, we state on the outset that for large-scale predictions, 
the overhead associated with computing more sophisticated
baseline properties can become non-negligible. 

Within $\Delta$-ML, property $p_1$ corresponding to some targetline level of theory 
is modeled as the sum of a closely related property $p_2$ (an observable which not necessarily results from the same operator as the one used to calculate $p_1$)
corresponding to an inexpensive baseline theory
and the usual ML contribution, i.e.~a linear combination of kernel-functions
\begin{equation}
p_1^{\rm target}({\bf d}_q^{\rm target}) \; \approx \; p_2^{\rm base} ( {\bf d}_q^{\rm base} ) 
+ \sum_{t=1}^N c_t k(|{\bf d}_q^{\rm base} - {\bf d}_t^{\rm base} |).
\label{eq:krr}
\end{equation}
Note that for predictions of targetline properties of new compounds 
the molecular structures computed at the baseline level of theory suffice. 
As such, the ML contribution accounts not only for the change in level of theory 
(i.e. vertical changes going from baseline to targetline in Fig.~\ref{fig:flowdelta}) or the
change in operator used to calculate the property,
but also for all changes due to structural differences between baseline and targetline structures. 
It is likely that this approach is successful for energies when such changes are small. 
Fortunately, molecular geometries are often already very well accounted for when using inexpensive semi-empirical quantum chemistry methods.

The regression problem within the $\Delta$-ML {\em Ansatz} corresponds simply to
\begin{equation}
({\bf K}+\lambda {\bf I})\ {\bf c} = {\bf p}_1^{\rm target}-{\bf p}_2^{\rm base}=
\Delta{\bf p}_{\rm base}^{\rm target}.
\label{eq:l2}
\end{equation}
Note that for ${\bf p}_2^{\rm base} =c$ $\Delta {\bf p}$ reduces to ${\bf p}_1^{\rm target}$, 
i.e.~we recover the ordinary ML problem of directly modeling a targetline property (shifted by $c$). 

The first $\Delta$-ML study, Ref.~\cite{DeltaPaper2015}, reports for several 
baseline, targetline combinations and demonstrates the feasibility to model
various molecular properties ranging from electronic energy, thermochemistry energetics, 
molecular entropy, as well as electron correlation energy. 
Using a trainingset size of 1\,k, the direct ML model, trained to model B3LYP heats of atomization, 
leads to a MAE of 13.5 kcal/mol for 133\,k out-of-sample predictions for molecules from QM9.
When relying on the widely used semi-empirical quantum chemistry model PM7, the MAE decreases to 7.2 kcal/mol.
Combining the two approaches into a 1\,k $\Delta_{\rm PM7}^{\rm B3LYP}$-ML model reduces the MAE to 4.8 kcal/mol. 
When increasing the data available to the ML model to 10\,k training examples, 
the MAE of the resulting 10\,k $\Delta_{\rm PM7}^{\rm B3LYP}$-ML model 
drops to 3 kcal/mol for the remaining 122\,k out-of-sample molecules. 
Such accuracy, when combined with the execution  speed of the baseline PM7, for
geometry relaxation, as well as calculation of atomization enthalpy, has enabled 
the screening of atomization enthalpies for 124\,k out-of-sample molecules at a level
of accuracy that is typical for B3LYP---within roughly 2 CPU weeks.
Note that the computational cost for obtaining the corresponding B3LYP atomization enthalpies
brute-force for the exact same molecules consumed roughly 15 CPU years of modern compute hardware.

\begin{figure*}
\centering
\includegraphics[width=12.5cm, angle=0.0, scale=1]{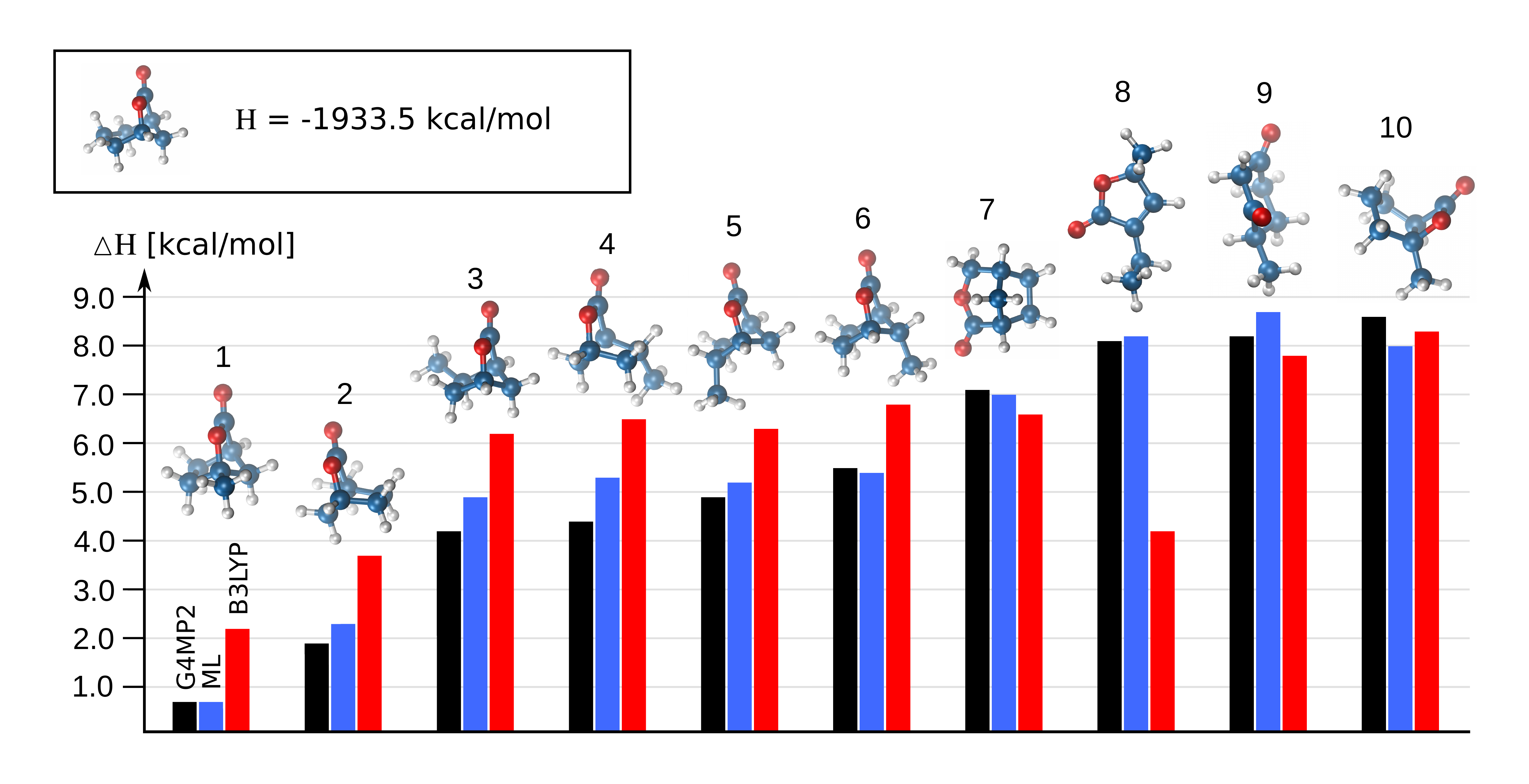}                               
\caption{
Calculated reaction enthalpies at 298.15 K between the most stable molecule with
C$_7$H$_{10}$O$_2$ stoichiometry (inset), 
and its ten energetically closest isomers in increasing order, according to targetline method G4MP2 (black). 
1\,k $\Delta_{\rm B3LYP}^{\rm G4MP2}$-ML model predictions are blue, B3LYP is red.
(Reprinted with permission from \cite{DeltaPaper2015}, Copyright (2015), 
ACS Publishing.)
}
\label{fig:isomerization}
\end{figure*}

In Ref.~\cite{DeltaPaper2015} the $\Delta$-ML approach has also been shown to reach chemical accuracy, i.e.~1 kcal/mol for atomization energies. 
More specifically, the performance of a 1\,k $\Delta_{\rm B3LYP}^{\rm G4MP2}$-ML model has been investigated for
predicting the stability ordering among 10\,k out-of-sample diastereomers generated
from all the constitutional isomers of C$_7$H$_{10}$O$_2$ present in the 134\,k dataset.
The 10 isomers closest in energy to the lowest lying isomer, predicted by $\Delta_{\rm B3LYP}^{\rm G4MP2}$-ML,
are shown together with validating G4MP2 and B3LYP predictions in Fig.~\ref{fig:isomerization}. 
We note that the $\Delta_{\rm B3LYP}^{\rm G4MP2}$-ML predictions typically agree with G4MP2 by less than 1 kcal/mol, 
a substantial improvement over the B3LYP predictions. 
Furthermore, the ML correction also rectifies the incorrect B3LYP ordering of the isomers (B3LYP predicts the eighth isomer to be lower in energy than isomers 3-7).
This latter point is crucial for studying competing structures such as they would occur under reactive conditions
in chemical reactions of more complex molecules.

Table~\ref{tab:deltaml} gives an overview for mean absolute out-of-sample prediction errors for 
1\,k $\Delta$-ML predictions of energetics in various datasets as well as for various baseline and targetline combinations.  
There is a general trend: The more sophisticated the baseline the smaller the prediction error. 
For predicting G4MP2 atomization heats, the pure density functional method PBE baseline results in a MAE $<$ 1 kcal/mol. 
The hybrid functional B3LYP quenches the error further down by 0.2 kcal/mol, 
but at the cost of the added overhead associated with the evaluation of the exact exchange terms. 
Such high accuracy might still be worth the while for reliably forecasting the stability ranks of competing isomerization reaction products,
which often require quantitative accuracy beyond that of DFT. 

\begin{table*}[htp]                                                                       
\caption{ 
Comparison of mean absolute errors (MAEs, in kcal/mol) for $\Delta$ML-predicted energetics ($E$) 
across datasets, and baseline/targetline combinations for a fixed trainingset size of 1\,k:  
$H$ is enthalpy of atomization;
$E_0$ is the energy of the electronic ground state;  
$E_j$ is $j^{\rm th}$ singlet-singlet transition energy;
{\em gap} refers to difference in the energies of HOMO and LUMO; 
{\em shifted} refers to constant shifts of -0.31 eV and 0.19 eV that have been added 
to TDPBE/def2SVP predicted $E_1$ for molecules with saturated $\sigma$
and unsaturated $\pi$-chromophores, respectively.
Zero baseline corresponds to directly modeling the targetline property.
For datasets with $N$ molecules, MAEs in predictions for
$N$-1\,k out-of-sample molecules are reported.
The last column indicates the reference where these numbers have been taken from.
        }                                                                                 
\label{tab:deltaml}                                                                           
\begin{tabular}{l   l   l   l   c }
\hline 
\hline 
\multicolumn{ 1}{l}{dataset~~~~~~~~~} &                                                           
\multicolumn{ 1}{l}{$E$~~} &                                                             
\multicolumn{ 1}{l}{baseline~~~~~~~~~~~~~~} &
\multicolumn{ 1}{l}{targetline~~~~~~} &
\multicolumn{ 1}{l}{MAE [kcal/mol]} \\[0.5ex]  
\hline 
  134\,k GDB-9           &         $H$     &     0  &        B3LYP/6-31G(2df,p)      &     13.5 ~~~ \cite{DeltaPaper2015} \\
                         &         $H$     &  PM7   &        B3LYP/6-31G(2df,p)      &     4.8  ~~~ \cite{DeltaPaper2015}  \\
                         &                 &        &                                &            \\
6\,k C$_7$H$_{10}$O$_2$  &         $H$     &     0  &        G4MP2                   &     5.1 ~~~  \cite{DeltaPaper2015} \\
                         &         $H$     &  PM7   &        G4MP2                   &     3.7 ~~~  \cite{DeltaPaper2015} \\
                         &         $H$     &  PBE/6-31G(2df,p)   &        G4MP2      &     0.9 ~~~  \cite{DeltaPaper2015}  \\  
                         &         $H$     &  B3LYP/6-31G(2df,p)   &        G4MP2    &     0.7 ~~~  \cite{DeltaPaper2015} \\
                                                  &                 &        &                                &            \\
6\,k C$_7$H$_{10}$O$_2$  &         $E_0$     &  HF/6-31G(d)    &        CCSD(T)/6-31G(d)        &     2.9 ~~~  \cite{DeltaPaper2015} \\
                         &         $E_0$     &  MP2/6-31G(d)   &        CCSD(T)/6-31G(d)        &     1.0 ~~~  \cite{DeltaPaper2015} \\
                         &         $E_0$     &  CCSD/6-31G(d)  &        CCSD(T)/6-31G(d)        &     0.4 ~~~  \cite{DeltaPaper2015} \\
                                                  &                 &        &                                &            \\
   22\,k GDB-8           &         $E_1$   &      0           &    CC2/def2TZVP      &    12.1 ~~~\cite{ML_TDDFTEnrico2015} \\
                         &         $E_1$   &   {\it gap}, TDPBE/def2SVP            &    CC2/def2TZVP      &    9.2 ~~~ \cite{ML_TDDFTEnrico2015}\\
                         &         $E_1$   &    $E_1$, TDPBE/def2SVP            &    CC2/def2TZVP      &    3.8 ~~~ \cite{ML_TDDFTEnrico2015}\\
                         &         $E_1$   &    $E_1$, TDPBE/def2SVP, {\em shifted}           &    CC2/def2TZVP      &    3.0 ~~~ \cite{ML_TDDFTEnrico2015}\\
                                                                           &                 &        &                                &            \\
    22\,k GDB-8           &         $E_2$   &    $E_2$, TDPBE/def2SVP            &    CC2/def2TZVP      &    5.5 ~~~ \cite{ML_TDDFTEnrico2015}\\
\hline 
\hline 
\end{tabular}
\end{table*}

\begin{figure*}
\centering
\includegraphics[width=8.8cm, angle=0.0, scale=1]{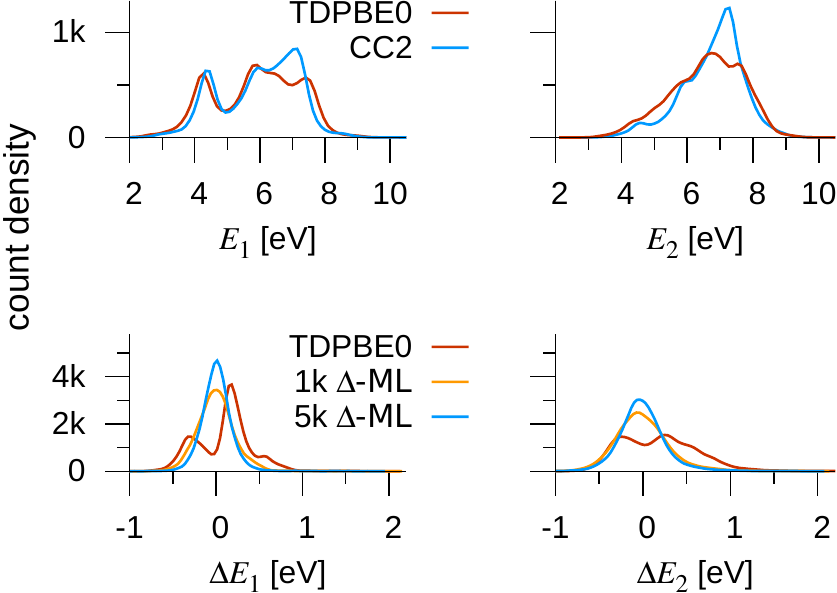}                               
\caption{
Density distributions of electronic excitation energies (top) and predicted errors (bottom). 
Top: Densities of first (left), and second (right) singlet-singlet transition energies 
of 17\,k organic molecules with up to eight CONF atoms, at the CC2/def2TZVP targetline, and
TDPBE0/def2SVP baseline levels of theory. 
Bottom: Densities of errors in predicting the values of $E_1$ and $E_2$ using $\Delta_{\rm TDPBE0}^{\rm CC2}$-ML models 
based on 1\,k (orange), and 5\,k (red) training molecules. 
(Reprinted with permission from \cite{ML_TDDFTEnrico2015}, Copyright (2015), AIP Publishing LLC.)
}
\label{fig:distri}
\end{figure*}

In a more recent study, the applicability of the $\Delta$-ML idea to the modeling of lowest 
electronic excitation energies at the CC2 level using TDDFT baseline values for 22\,k GDB-8 molecules has also been investigated~\cite{ML_TDDFTEnrico2015}. 
Results from that are reported in Table~\ref{tab:deltaml}. 
While comparing the performance of 1\,k models for the lowest two excitation energies, $E_1$, and $E_2$, one notes that the prediction is slightly worse for $E_2$. 
Both $\Delta$-ML studies \cite{DeltaPaper2015,ML_TDDFTEnrico2015} found the distribution
of errors in the ML predictions to be unimodal, {\em irrespective} of the nature of the distribution of the deviation between the targetline and baseline. 
This point is illustrated in Fig.~\ref{fig:distri} for the modeling of $E_1$, and $E_2$ for the 22\,k GDB-8 set. 
Ref.~\cite{ML_TDDFTEnrico2015} also showed how systematic shifts in target property,
can be included explicitly in the baseline to improve ML prediction accuracy. 

\section{Atoms in Molecules}
A number of molecular observables manifest themselves as properties of atoms perturbed by the local chemical environment in molecules. 
Common examples include NMR chemical shifts, and core-level ionization energies. 
Some interesting atomic properties that are more difficult to measure experimentally include 
partial charges on atoms, forces on atoms for MD simulation,
components of tensor properties such as dipole and quadruple moments, atomic polarizability, etc. 
To directly model atomic forces, as opposed to first modeling the energies over suitably sampled 
molecular structures and subsequent differentiation, has the advantage that the oscillations between training
data points, inherent to any statistical machine learning model, are not amplified through differentiation.
Also, the generation of force training data does not imply additional investment in terms of {\it ab initio} calculations, 
because of the availability of analytic derivatives by virtue of the Hellmann-Feynman theorem~\cite{HF}: 
The evaluation of atomic forces for all atoms in a molecule typically incurs computational efforts similar to 
calculating the total potential energy of just one molecule---at least within mean-field theories such as DFT. 
But even at the levels of correlated wavefunction methods such as CCSD and QCISD(T), 
the cost of computing analytic derivatives is only 4--5 times higher than that of computing a 
single point energy in the case of small molecules \cite{RaghuRauhut2015}.

One of the challenges in the modeling of properties of atoms in molecules consists of finding a suitable local chemical descriptor. 
For vector valued atomic properties, an additional challenge is the choice of an appropriate reference frame. 
After resolving these issues, the predictive accuracy of ML models trained on sufficiently large data sets
are transferable to similar atomic environments, even when placed in larger systems. 
In Ref.~\cite{MLatoms_2015}, such ML models have been introduced for NMR chemical shifts of carbon and hydrogen ($^{13}$C~$\sigma$, $^{1}$H~$\sigma$),
for core-level ionization energies of C atoms, and for atomic forces. 
All atomic properties used for training and tested were calculated at the PBE0/def2TZVP level for many thousands of organic GDB molecules~\cite{GDB17}. 
For the forces, a principal axis system centered on the query atom was chosen as reference frame, 
and three independent ML models were trained, one for each element of the atomic forces 
projected into this local coordinate system. 
While in many quantum chemistry codes the molecular geometry is oriented along the 
principal axis system by default, popularly known as {\tt standard orientation}, for the purpose of learning atomic forces 
reference frames have to be found for each atom independently. 
Cartesian coordinates of the query atom as well as its neighboring atoms within a certain cut-off radius
and sorted by their distances to the query atom were also oriented according to the same reference frame. 
A localized atomic variant of the CM was used with
first row/column corresponding to the query atom, and with indices of all other atoms
being sorted by their distance to the query atom. 
Furthermore, a cut-off radius was optimized for each atomic property --- 
atoms that lie beyond this cut-off do not feature in the atomic CM. 
Non-equilibrium geometries, necessary to compute non-zero training and testing forces, 
were sampled via stochastic displacements along normal modes.
For computing NMR chemical shifts tetramethylsilane (TMS) was chosen as the reference compound,
and shifts in C (1$s$) core-level energies were considered using methane as the reference. 

All aforementioned local atomic properties were modeled using Laplacian kernel and Manhattan distance. 
When trained on 9\,k randomly sampled atoms present in 9\,k random molecules from the 134\,k QM9 set, 
the resulting models were validated with out-of-sample predictions of the same properties for 
1\,k atoms drawn at random from the remaining 125\,k  molecules not part of training. 
The accuracy achieved reached the order of magnitude of the reference DFT method.
Specifically, a relative mean square error (RMSE) of 4 units was obtained for the properties of C atoms 
($^{13}$C~$\sigma$ (ppm), C 1$s$~$\sigma$
(ppm), $F_C$ m.a.u.),
while for the force on H atoms, $F_H$, an RMSE of 0.9 was obtained. 
For $^{1}$H~$\sigma$ with a property range of only 10 ppm, an RMSE of 0.1 ppm has been reported. 
Very similar errors were noted when {\it trained} on
short polymers made from the GDB-9 molecules, and {\it tested} on longer saturated 
polymers of length up to 36 nm. 
Applicability of the local ML model across CCS was demonstrated by accurate predictions of $^{13}$C~$\sigma$-values 
for all the 847\,k C atoms present in the 134\,k molecular set.

Refs.~\cite{MLOTF_2015,MLQMMMSupercompSandro_2015,RampiForces_2015} also describe ML models of atomic forces which can be 
used to speed up geometry relaxation or molecular dynamics calculations.
Note that these forces are always being modeled within the same chemical system. 
In other words, they can neither be trained nor applied universally throughout chemical space, 
and they inherently lack the universality of the ML models reviewed here within. 
In Ref.~\cite{MTP_Tristan2015}, atomic ML models are introduced which can be used as parameters in universal FFs, i.e.~throughout chemical space. 
More specifically, transferable ML models of atomic charges, dipole-moments, and multipole-moments were trained and used 
as dynamically and chemically adaptive parameters in otherwise fixed parametrized FF calculations. 
The intermolecular interactions accounted for through use of such dynamic electrostatics, effectively accounting
for induced moments (or polarizable charges), in combination with simple estimates of many-body dispersion~\cite{Voronoi_Tristan2014},
resulted in much improved predictions of properties of new molecules.  
In Ref.~\cite{Dral_2015} an atomic parameter, used within semi-empirical quantum chemistry,
has been improved through use of ML models trained in chemical space. 

The inherent capability of atomic property models to scale to larger molecules
provides another argument why building a ML model of atomic forces is advantageous over building 
a ML model of energies with subsequent differentiation---the assumptions about how to partition the system are more severe in case of the latter.
As such, while an atomic ML model will not be able to extrapolate to entirely new chemical environments that were not part of its training training set, 
it is well capable of being used for the prediction of an arbitrarily large diversity of macro- and supra-molecular structures
such as polymers or complex crystal phases---as long as the local chemical environments are similar. 
This point has been investigated and confirmed in Ref.~\cite{MLatoms_2015}:
For up to 35 nm (over 2\,k electrons), the error is shown to be independent of system size.
The computational cost remains negligible for such systems, while conventional DFT based calculations
of these properties require multiple CPU days.

\section{Crystals}
Various attempts have already been made to apply ML to the prediction of crystal properties. 
In Ref.~\cite{GrossMLCrystals2014}, a radial distribution function based descriptor 
is proposed that leads to ML learning models which interpolates between pre-defined atom-pairs. 
Ref.~\cite{ML4Crystals_Wolverton2014} exploits chemical informatics based concepts for the design of crystalline
descriptors which require knowledge of previously calculated QM properties, such as the band-gap. 
Ghiringhelli et al.~report on the design of new descriptors for binary crystals in Ref.~\cite{GhiringhelliSchefflerDescriptor_PRL2015}.
Through genetic programming they solve the inverse question of which atomic properties should be combined how in order to define a good descriptor. 
A periodic adaptation of the CM, as well as other variants, while leading to learning
rates that can systematically be improved upon, did lead to reasonable ML models for large training sets~\cite{MLcrystals_Felix2015}.
Very recently, however, a superior descriptor for ML models of properties of pristine crystals has been proposed~\cite{Elpasolite_2015}.
This descriptor encodes the atomic population on each Wyckoff-symmetry site for a given crystal structure.
As such, the resulting ML model is restricted to all those crystals which are part of the same symmetry group.
All chemistries within that group, however, can be accounted for. 
At this point one should note that already for binary cyrstals
the number of possible chemistries dramatically exceeds the number of possible crystal symmetries,
not mentioning ternaries or solid mixtures with more components.
The applicability of the crystal ML model has been probed by screening DFT formation energies of all the 2 M Elpasolite (ABC$_2$D$_6$) crystals
one can possibly think of using all main-group elements up to Bi.
Using an ML model trained on only 10\,k DFT energies,
the estimated mean absolute deviation of the out-of-sample ML prediction from DFT formation energies is less than 0.1 eV/atom.
When compared to experiment or quantum Monte Carlo results, such level of predictive accuracy corresponds 
roughly to what one would also expect for solids from generalized gradient approximated DFT estimates. 

\section{Conclusions and outlook}
\label{sec:Conclusions}
We have reviewed recent contributions which combine statistical learning with quantum mechanical concepts 
for the automatic generation of ML models of materials and molecular properties.
All studies reviewed have in common that after training on a given data set, 
ML models are obtained that are transferable in the sense that they can infer solutions 
for arbitrarily many (CCS is unfathomably large as molecular size increases) 
out-of-sample instances with an accuracy similar to the one of the deductive 
reference method used to generate the training set---for a very small, if not negligible, computational overhead. 
Note that the genesis of the training data is irrelevant for this {\em Ansatz}:
While all studies rely on pre-calculated results obtained from legacy quantum chemistry approximations 
which typically require substantial CPU time investments, training data could just as well have come from experimental measurements.
Depending on the reliability and uncertainty of the experimental results, however, we would expect the 
noise-parameter $\lambda$ (see Eq.~\ref{eq:regression}), to become significant in such a set-up.
Training-data obtained from computational simulation does not suffer from such uncertainties 
(apart from the noise possibly incurred due to lack of numerical precision), 
resulting in practically noise-free data sets. 

Comparing performance for molecules, nano-ribbon models, atoms, or crystals,
one typically observes satisfying semi-quantitative accuracy for ML models trained on at least one thousand training instances. 
Accuracies competitive with current state-of-the-art models which can reach or exceed experimental accuracy, 
are typically found only after training on many thousands of examples.
The main appeal of the ML approach reviewed is due to its simplicity and computational efficiency: 
After training, the
evaluation of the ML model for a new query instance according to Eq.~(1.1) 
typically consumes only fractions of a CPU second, while solving SE in order to evaluate a quantum mechanical observable easily requires multiple CPU hours.
As such, this approach represents a promising complementary assistance when it comes to bruteforce high-throughput quantum mechanics based screeening 
of new molecules and materials.

In summary, the reviewed studies, as well as previously published Refs~\cite{RuppPRL2012,Montavon_NIPS2012,Montavon2013,AssessmentMLJCTC2013},
report on successful applications of supervised learning concepts to the automatic generation of analytic materials and molecular
property models using large numbers of parameters and flexible basis functions.
Quantum properties considered so far include 
\begin{itemize}
\item[] atomization energies and thermochemical properties such as atomization enthalpies and entropies with chemical accuracy~\cite{DeltaPaper2015};
\item[] molecular electronic properties such as electron correlation energy, frontier orbital eigenvalues (HOMO and LUMO), electronic spread, dipole-moments, polarizabilities, excitation energies~\cite{DeltaPaper2015,ML_TDDFTEnrico2015,SingleKernel2015,BobPaper};
\item[] atomic properties such as charges, dipole moments, and quadrupole moments, nuclear chemical shifts, core electron excitations, 
atomic forces~\cite{MLatoms_2015,MTP_Tristan2015};
\item[] formation energies of crystals~\cite{MLcrystals_Felix2015,Elpasolite_2015};
\item[] electron transport as well as hybridization functions~\cite{QuantumTransportML2014,LF_DMFT_ML2014,LF_DMFT_ML2015}.
\end{itemize}
Performance is always being measured out-of-sample, i.e., for ``unseen'' molecules not part of training, and at negligible computational cost.
As such, we note that there is overwhelming numerical evidence suggesting that such ML approaches can be used as
accurate surrogate models for predicting important quantum mechanical properties with sufficient accuracy,
yet unprecedented computational efficiency--provided sufficient training data is available.
For all studies, data requisition has resulted through use of well established and standardized quantum chemical 
atomistic simulation protocols on modern super computers. 
This has also been exemplified by generating a consistent set of quantum mechanical structures and
properties of 134\,k organic molecules~\cite{DataPaper2014} drawn from the GDB-17 dataset~\cite{GDB17}.
The more conceptual role of representations and kernels, 
studied in Refs~\cite{FourierDescriptor,BobPaper,SingleKernel2015}, have been reviewed as well. 

The limitations of ML models lie in their inherent lack of transferability to new regimes.
Inaccurate out-of-sample predictions of properties of molecules or materials which differ substantially from the training set should therefore be expected. 
Fortunately, however, most relevant properties and chemistries are quite limited in terms of 
stoichiometries (at most $\sim$100 different atom types) and geometries 
(typically, for semi-conductors interatomic distances larger than 15{\AA} will hardly contribute 
to partitioned properties such as atomic charges or interatomic bonding; while interatomic distances smaller than 0.5{\AA} are exotic).
Furthermore, ``extrapolation'' is not a verdict but can sometimes be circumvented. 
As discussed above, due to the localized nature of some observables, 
atomic ML models can be predictive even if the atom is part of a substantially larger molecule. 
It has been demonstrated that these property models will be transferable to larger molecules containing similar local chemical environments, 
and that they scale with system size---as long as the underlying roperty is sufficiently well localized. 

Future work will show if ML models can also be constructed for other challenging many-body problems including vibrational frequencies,
reaction barriers and rate constants, hyperpolarizabilities, solvation, free energy surfaces, electronic specrat, Rydberg states etc.
Open methodological questions include optimal representations and kernel functions,
as well as the selection bias present in most (if not all) databases.

\begin{acknowledgments}
The writing of this review was funded by the Swiss National Science foundation (No.~PP00P2\_138932).
\end{acknowledgments}

\bibliography{literatur}{}
\bibliographystyle{apsrev-revcompchem}
\end{document}